\documentclass{aa}  

\usepackage{graphicx}
\usepackage{multirow}
\usepackage{xcolor}
\usepackage{txfonts}
\begin{document} 

   \title{On the seismic modelling of subgiant stars: testing different grid interpolation methods}

   \author{M. Clara \inst{1,2} \and
        M. S. Cunha \inst{1,3} \and
        P. P. Avelino \inst{1,2,3} \and
        T. L. Campante \inst{1,2} \and
        S. Deheuvels \inst{4} \and
        D. R. Reese \inst{5}}

   \institute{Instituto de Astrofísica e Ciências do Espaço, Universidade do Porto, CAUP, Rua das Estrelas, 4150-762 Porto, Portugal\\
            \email{miguel.clara@astro.up.pt}
        \and
             Departamento de Física e Astronomia, Faculdade de Ciências da Universidade do Porto, Rua do Campo Alegre, s/n, 4169-007 Porto, Portugal
        \and
             Université Côte d’Azur, Observatoire de la Côte d’Azur, CNRS, Laboratoire Lagrange, France
        \and
             IRAP, Université de Toulouse, CNRS, CNES, UPS, 14 avenue Edouard Belin, 31400 Toulouse, France
        \and
             LIRA, Observatoire de Paris, Université PSL, Sorbonne Université, Université Paris Cité, CY Cergy Paris Université, CNRS, 92190 Meudon, France}

   \date{Received 25/04/2024; Accepted: 19/12/2024}
 
  \abstract
   {Grid-based modelling techniques have enabled the determination of stellar properties with great precision.
   The emergence of mixed modes during the subgiant phase, whose frequencies are characterized by a fast evolution with age, can potentially enable a precise determination of stellar properties, a key goal for future missions like PLATO. However, current modelling techniques often consider grids that lack the resolution to properly account for the fast mode frequency evolution, consequently requiring the use of interpolation algorithms to cover the parameter space in between the grid models when applying model-data comparison methods.}
   {We aim at reproducing the $\ell$=1 mode frequencies within the accuracy limits associated with the typical observational errors ($\sim$0.1 $\mu$Hz) through interpolation on a grid of subgiant models.}
   {With that aim, we used variations of a two-step interpolation algorithm, which considered linear and cubic splines interpolation methods and different age proxies (physical age, scaled age, and central density).}
   {The best results were obtained using an algorithm that considers cubic splines interpolation along tracks, linear interpolation across tracks, and central density $\rho_\text{c}$ as the age proxy. This combination yielded, on average, an absolute error of 0.14 $\mu$Hz, but reached maximum absolute errors on the interpolated frequencies of 1.2 $\mu$Hz for some models, which is an order of magnitude higher than the typical observational errors. Furthermore, we investigated the impact on the accuracy of the interpolation from changes in the physical properties of the stars, showing, in particular, how the addition of core overshoot can affect significantly the interpolation results.}
   {}

   \maketitle


\section{Introduction}
\label{Sec.Introduction_Section}

As solar-type stars end their longest stage of evolution, by finishing the burning of hydrogen in their cores and leaving the main-sequence (MS) towards more evolved stages of evolution, their physical and chemical structure allow the emergence of new interesting features.
Supported by a combination of stellar pulsation theory and a tremendous amount of high-quality asteroseismic data from space missions like CoRoT (Convection, Rotation and planetary Transits; \cite{COROT_Mission_2006}), Kepler \citep{Kepler_Gilliland_2010, Kepler_Koch_2010}, TESS (Transiting Exoplanet Survey Satellite; \cite{TESS_Ricker_2015, TESS_Campante_2016, TESS_Cunha2019}), and the upcoming PLATO mission (PLAnetary Transits and Oscillations of stars; \cite{Plato_Mission_2014, PLATO_Cunha2021, Rauer2023, Goupil2024}), these features can be used to make robust inferences on stellar properties, internal structure and dynamics.

During the MS, turbulent motions drive global pulsation modes: the stochastic oscillations of solar-like pulsators. These are acoustic (or p-) modes and are maintained by the gradient of pressure fluctuation, being mostly sensitive to conditions in the outer part of the star. This is so because their propagation cavities are located between the surface and an inner turning point (whose depth is strongly dependent on the mode degree $\ell$). 
In addition, it is possible that MS solar-like stars also support internal gravity waves maintained by gravity acting on density fluctuations, essentially trapped beneath the convective envelope in a cavity that is mostly independent of mode degree \citep{Unno_book_1989, Aerts_ChristensenDalsgaard_book_2010, Cunha2018}. Nevertheless, despite some claims \citep{Garcia2010, Fossat2017}, it is not yet clear that these gravity (or g-) modes have been observed in the Sun.

The properties of p- and g- modes depend directly on the stellar structure, hence varying significantly with stellar mass and evolution. As a star evolves beyond the MS into the so-called subgiant phase, hydrogen fusion in a shell outside of the core supplies the star with energy as its core contracts and the buoyancy frequency increases significantly towards the centre. These structural changes cause the frequency of the g-modes to approach the frequency of the p-modes, resulting in the appearance of mixed modes. These modes are maintained by gravity acting on density perturbations in the deep interior, and by the gradient of the pressure perturbation in the outer layers, strongly constraining the interior structure of the star \citep{DeheuvelsMichel2011}. They are also characterized by frequencies that have a fast evolution with age, which can potentially be used to determine stellar properties with great precision \citep{Li2019}.

Despite their potential, subgiant stars are not so well-studied as both MS stars and red giants, since they are both less common than MS stars and fainter than red giants. However, major advances are expected as subgiants constitute a significant fraction of the asteroseismic targets of TESS and of the upcoming PLATO mission.

Several fast, stable and robust optimization algorithms are now available to the community, such as AIMS (Asteroseismic Inference on a Massive Scale, \cite{Reese_AIMS_2019}), BASTA (BAyesian STellar Algorithm, \cite{Aguirre2022}), SPInS (Stellar Parameters INferred Systematically, \cite{Reese_Lebreton_SPINS_2020}), and PARAM \citep{Rodrigues2017}. 
These algorithms provide a powerful diagnostic tool for the determination of stellar properties, that allow the determination of a representative set of models that best reproduce a given set of classical and asteroseismic constraints. 

Most of the optimization algorithms are grid-based, requiring access to pre-computed grids of stellar models that span a given parameter space.
This is quite different from a so-called on-the-fly modelling approach where the goodness of the fit is calculated while generating the stellar models. In this method, the set of models generated is of use only for the target being modelled, with the generation of a new set of models being required each time a new target is considered. This makes the on-the-fly modelling computationally expensive and less appropriate for a pipeline of an instrument observing a vast number of stars.
However, there is also a drawback associated with the grid-based method: when increasing the number of parameters of the models, the number of models on the stellar grid also increases, increasing the required storage and possibly decreasing the computational efficiency. Thus, expanding the grid to include more input parameters generally requires reducing the resolution of the initial grid which, in turn, reduces the precision with which stellar properties are inferred. 
Note that, despite their differences, these two modelling approaches are not necessarily mutually exclusive, as the result of grid searches may be used to restrict the region of parameter space being considered for the more refined model-on-the-fly analysis. 

In order to guarantee a given precision on the inferred stellar properties needed, e.g., to ensure that the science requirements of a mission are satisfied, it may be required to couple a grid-based approach with an interpolation algorithm. These algorithms are used to compensate for the lack of resolution in model grids, by providing an approximate description of model predictions (e.g., the global and seismic properties) at points of the parameter space located in between the models of the initial pre-computed discrete grid.

Current optimization codes, like AIMS and BASTA, use (or may use) interpolation algorithms.
Typically, the interpolation procedure considers a 2-step process: (i) interpolation along the tracks, according to a chosen age proxy
; and (ii) interpolation between evolutionary tracks, which amounts to interpolation in the parameter space defined by the grid parameters, excluding age.

\cite{Reese_AIMS_2019} discuss the results of interpolation tests conducted while fitting observational data with AIMS. The authors compare their results with the typical uncertainties found in the literature for a MS target, 16 Cyg A, whose smallest and average uncertainties for $\ell$=0 modes are 0.04 and 0.08 $\mu$Hz, respectively, and for a red giant branch (RGB) target, KIC 4448777, whose smallest and average uncertainties for $\ell$=0 modes are 0.014 and 0.018 $\mu$Hz, respectively.
For interpolation along evolutionary tracks, interpolation errors for $\ell$=0 modes were found to be smaller than both the smallest and average uncertainties from literature values, except for a small fraction of RGB tracks (less than 3\% of the number of RGB tracks) characterized by low masses and high metallicities.
For cross-track interpolation, their root mean square average errors are comparable to their literature reference value of 0.08 $\mu$Hz.

For BASTA, \cite{Aguirre2022} only discuss the results of the complete interpolation procedure and restrict the analysis to MS stars. Using an interpolation procedure similar to that employed in AIMS, they show that the fractional differences in $\Delta\nu$ between the original and the interpolated track are at the $10^{-4}$ level, an order of magnitude smaller than the relative uncertainty in $\Delta\nu$ measured for the best Kepler targets.
\cite{Li2020b} test the interpolation of $\ell$=1,2 oscillation mode frequencies while investigating the dependencies of seismic stellar ages for 36 Kepler subgiants on helium fraction, the mixing-length parameter and metallicity.
Using cubic splines interpolation and physical age, $\tau$, as the age proxy, the authors perform interpolation along a mass-track with 1.20 M$_\odot$, where models to be interpolated are separated by a time step of $6\times10^6$ yr.
For $\ell$=1 modes, the differences between interpolated frequencies and their true values were mostly below $\sim$0.05 $\mu$Hz, especially for frequencies in the range of 0.5-1.5 $\nu_\text{max}$.
Using the same method, interpolation of $\ell$=2 mode frequencies was found to work well for p-dominated modes, but not for g-dominated ones. For 85\% of the tested modes, frequency differences were found to be smaller than 0.2 $\mu$Hz, but, for the remaining poorly fitted g-dominated modes, differences can go up to 0.5 $\mu$Hz.

In summary, among the literature works concerning grid-based modelling of solar-like pulsators, only \cite{Li2020b} performed interpolation tests on grids of subgiant stars. Yet, even in that case, the work focused exclusively on interpolation along evolutionary tracks. This highlights a lack of quantification of the accuracy of interpolation algorithms in the subgiant evolutionary phase.

In this work, we perform a characterization of classical interpolation algorithms applied to the determination of the pulsation properties in the subgiant evolutionary phase, and discuss the problems associated with them, establishing which method is capable of most accurately describing model frequencies interpolated both along and across evolutionary tracks of a two-parameter stellar grid. As a reference, we compare the interpolation errors with typical observational errors of $\delta\nu \sim$0.1$\mu$Hz \citep{Li2020b} that we set as our accuracy aim.

In Section \ref{Sec.Impact_of_Mixed_Modes_in_Subgiants}, we discuss what makes interpolation during the subgiant phase particularly challenging, further motivating our work.
In Section \ref{Sec.Characterization_of_the_Method} we describe the grid of stellar models and the classical interpolation methods we intend to characterize.
In Section \ref{Sec.Characterization_of_the_Interpolation_Errors}, we analyse the results and the interpolation errors associated with the application of the various interpolation methods to the stellar grid, while considering different interpolation methods and age proxies. We start by considering only interpolation along an evolutionary track, before addressing the more general case of interpolation across the mass tracks of the grid. We also discuss the influence that different grid resolutions have on the interpolation errors.
In Section \ref{Sec.Influence_of_Core_Properties_in_Interpolation}, we investigate the nature of a problem caused by interpolating mass-tracks across the mass region where MS models start developing a convective core, discussing how the overshoot parameter $f_\text{ov}$ influences the results. Finally, a summary of our main results is presented in Section \ref{Sec.Conclusions_Article}.


\section{The impact of subgiant mixed modes in interpolation}
\label{Sec.Impact_of_Mixed_Modes_in_Subgiants}

There have been numerous theoretical studies stating the potential of using the avoided crossings to accurately determining stellar ages during the subgiant phase, but none has fully considered their impact in grid interpolation.
In this section, we motivate our work by discussing why some age proxies are more adequate than others when applying interpolation to typical grids of stellar subgiant models.

\begin{figure}
  \resizebox{\hsize}{!}{\includegraphics{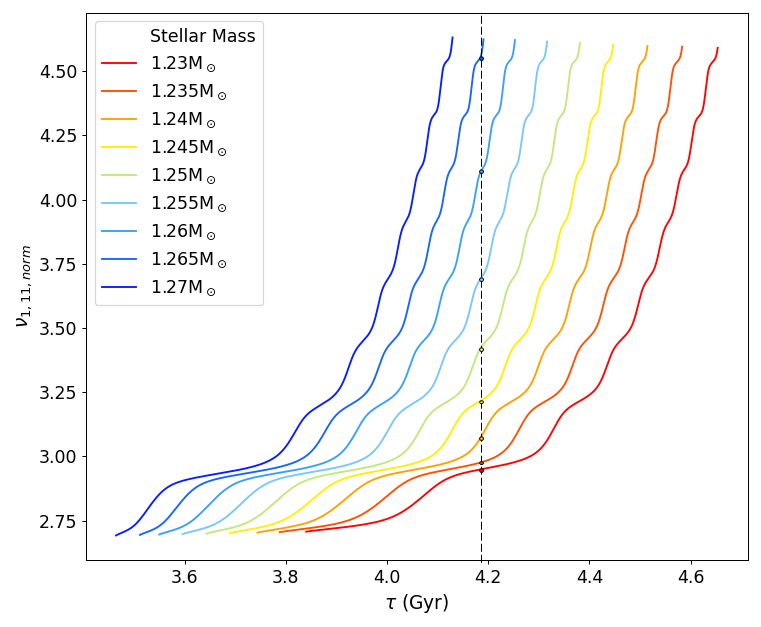}}
  \resizebox{\hsize}{!}{\includegraphics{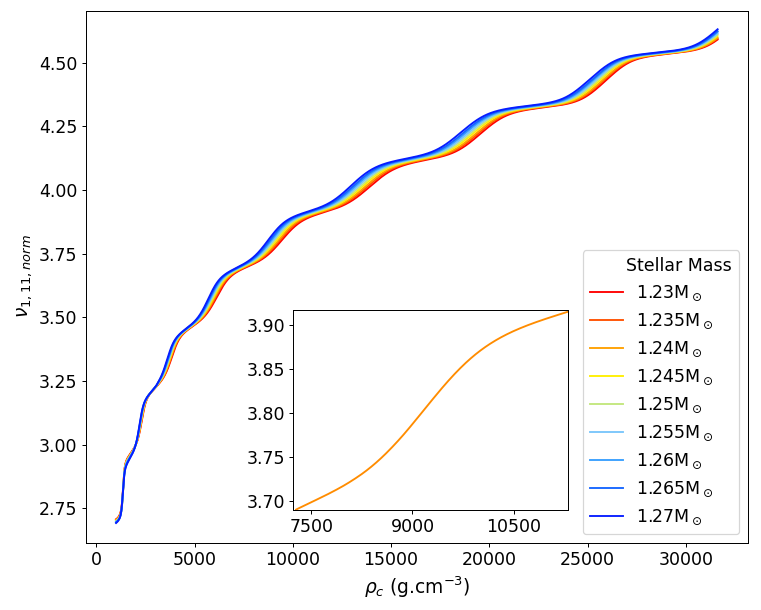}}
  \caption{Evolution of the normalized frequencies of non-radial $\ell$=1 modes, with radial order $n$=11, for evolutionary tracks of a stellar grid (the details of which will be given in Section \ref{Sec.Characterization_of_the_Method}) with masses between 1.23 and 1.27 M$_\odot$, as a function of the physical age $\tau$ (top) and the central density $\rho_\text{c}$ (bottom). The points connected by a dashed line in the top panel indicate models with an age of 4.18 Gyr. The bottom panel displays a subplot that zooms on a particular avoided crossing for the mass $M$=1.24 M$_\odot$.}
  \label{Fig.Phase_Comparison_of_ell1_Modes_as_Function_of_Physical_Age}
\end{figure}

Along a given evolutionary track, p-mode oscillation frequencies change following the evolution of the mean density. In order to remove the impact of that evolutionary effect, we considered normalized frequencies defined by
\begin{equation}
    \nu_{\ell,n,\text{norm}} = \nu_{\ell,n}/\sqrt{GM/R^3},
    \label{Eq.Normalization_of_Frequencies}
\end{equation}
where $G$ is the gravitational constant, $M$ is the stellar mass, and $R$ is the stellar radius of the corresponding model.

During the subgiant phase, we see the emergence of mixed modes whose frequencies exhibit a fast evolution, especially near the avoided crossings, that last about a couple tens of Myrs. For a fixed age, the same mode might be more acoustic- or gravity-dominated for different evolutionary tracks. 
This is shown in the top panel of Figure \ref{Fig.Phase_Comparison_of_ell1_Modes_as_Function_of_Physical_Age}, which displays the evolution of the normalized frequency of a dipole mode of radial order $n$=11 over various mass-tracks, where we highlight a fixed value of age, of 4.18 Gyr, with a vertical dashed line. Because of this difference in the mode evolution at fixed age, when interpolating between different evolutionary tracks, we may be combining a mode that is evolving essentially as if it were acoustic (like in the case of the mass-track with 1.23 M$_\odot$), with a mode that is going through an avoided crossing (like in the case of the mass-track with 1.24 M$_\odot$).
This difference in the mode evolution phase at fixed age is expected to impact the accuracy of the interpolation. One possible way to improve on that is to use proxies of age that better align the avoided crossings of adjacent mass-tracks. One such proxy is the central density, $\rho_\text{c}$ as illustrated in Figure \ref{Fig.Phase_Comparison_of_ell1_Modes_as_Function_of_Physical_Age}.
From the bottom panel, we see that avoided crossings appear more aligned when displayed as a function of the central density, $\rho_\text{c}$.

To understand why $\rho_\text{c}$ may be a better age proxy, we consider the evolution of the properties of the stellar core during the subgiant phase. When the H-burning ceases from being the main process of energy production in the core, the regions surrounding the isothermal He-core contract in order to increase pressure and temperature, triggering the H-burning process in a shell-boundary region around the core.
This change in the structure influences the profile of the Brunt-Väisälä frequency in the stellar core. The increase in the Brunt-Väisälä frequency causes the frequency of g-modes to approach the frequency of p-modes, giving rise to a series of avoided crossings. In a fully ionized environment, the Brunt-Väisälä frequency can be written as 
\begin{equation}
    N^2 = \frac{g}{H_P}(\nabla_\text{ad}-\nabla)+\frac{g}{H_P}(\nabla_\mu),
\end{equation}
where $g$ is the local gravitational acceleration inside the star, $H_P$ is the local pressure scale height, $\nabla_\text{ad}$ is the adiabatic gradient, $\nabla$ is the temperature gradient, and $\nabla_\mu$ is the gradient of the mean molecular weight.
Near the centre ($r\rightarrow0$), the local gravity inside the compact radiative subgiant core can be written as $g(r)\sim G\rho_c r$. For an ideal gas in hydrostatic equilibrium, we thus find that the Brunt-Väisälä frequency scales as
\begin{equation}
    N^2 \sim \frac{g(r)}{H_\text{P}} \sim \frac{\rho_\text{c}^3}{P_\text{c}}r^2 \sim \frac{\rho_\text{c}^2 \mu_\text{c}}{T_\text{c}}r^2.
    \label{Eq.Brunt_Vaisala_Frequency_in_Subgiant_Cores}
\end{equation}

In early post-MS models, the central temperature, $T_\text{c}$, and the central mean molecular weight, $\mu_\text{c}$, hardly change in the isothermal core where no nuclear reactions are happening.
This leaves $\rho_\text{c}$ as the main dependency of the Brunt-Väisälä frequency scaling in Equation \ref{Eq.Brunt_Vaisala_Frequency_in_Subgiant_Cores}. $\rho_\text{c}$ increases monotonically during the subgiant phase due to the contraction of the inner regions of the star.
It can, therefore, be considered as an age proxy for interpolation, with potential advantages compared to using physical age. In the following sections, we will consider $\rho_\text{c}$ and other age proxies to test grid interpolation accuracy during the subgiant phase.


\section{Characterization of the method}
\label{Sec.Characterization_of_the_Method}

We start by describing the grid of stellar models used for our simulation tests and the procedure adopted to characterize the accuracy of linear and higher-order interpolation algorithms. 

\subsection{Input physics of the stellar grid}
\label{Sec.Input_Physics_of_the_Stellar_Grid}

To perform tests that characterize the quality of interpolation, we compute a mass-age grid of evolutionary tracks in the subgiant phase. To this end, we used Modules for Experiments in Stellar Astrophysics (MESA, version r12778), an open-source stellar evolution package that is undergoing active development \citep{Paxton2011, Paxton2013, Paxton2015, Paxton2018, Paxton2019}.
We adopted the solar composition from \cite{Asplund_Grevesse_Sauval_2009} and the OPAL opacity tables given by \cite{Grevesse_Sauval1998}, which were supplemented by the low-temperature opacities from \cite{Ferguson2005}. 
We used the nuclear reactions rates obtained from the Joint Institute for Nuclear Astrophysics Reaction Library (JINA REACLIB, \cite{Cyburt2003}) version 2.2, with specific rates for $^{12}\text{C}(\alpha,\gamma)^{16}\text{O}$ \citep{Kunz2002} and $^{14}\text{N}(p,\gamma)^{15}\text{O}$ \citep{Imbriani2005}.
The boundary conditions at the stellar surface were established using the Krishna-Swamy atmosphere \citep{Krishna_Swamy1966}.

Our aim with this work is to test the interpolation across evolutionary tracks of different masses. Even though the ultimate goal is to test interpolation in a full grid where all parameters vary, we decided to keep things as simple as possible, by fixing all the physical inputs of the grid except for mass. Thus, we fixed metallicity at [Fe/H] = -0.1 dex, and varied the stellar mass between values of 1.150 and 1.350 M$_\odot$, in steps of 0.005 M$_\odot$. 
The goal was to produce a small but dense grid of stellar models in the subgiant phase, with masses crossing the region where small convective cores develop during the main sequence.
The initial helium fraction was determined following the law of Galactic enrichment,
\begin{equation}
    Y_\text{i} = Y_0 + \frac{\Delta Y}{\Delta Z} Z_\text{i},
\end{equation}
where the primordial helium abundance is $Y_0$=0.2485, according to \cite{PlanckCollaboration2016}, and the ratio $\Delta Y/\Delta Z$ is estimated to be 1.36 \citep{Balser2006, Nsamba2021}.
We also adopted the solar chemical mixture provided by \cite{Asplund_Grevesse_Sauval_2009}, $(Z/X)_\odot$=0.018099. 
The mixing-length parameter (first established in \cite{Bohm_Vitense1958}) is defined by $\alpha_\text{MLT}=\ell_\text{MLT}/H_\text{P}$, where $\ell_\text{MLT}$ is the mixing length, and $H_\text{P}$ is the pressure scale height. In MESA, $\alpha_\text{MLT}$ is implemented by the \texttt{mlt} module for convection, as presented by \cite{Cox_Giuli1968}, and, for this grid, was fixed as 1.711.
Convective overshooting was implemented via an exponential decay of the convective diffusion coefficient beyond the convective boundary, according to \cite{Herwig2000},
\begin{equation}
    D_\text{ov}=D_{\text{conv,0}}.\text{exp}\left(-\frac{2z}{f_\text{ov}H_\text{P}}\right),
    \label{Eq.Convective_Overshoot_Parameter}
\end{equation}
where $D_{\text{conv,0}}$ is the diffusion coefficient at the convective border, $z$ is the distance from the boundary of the convective region, and $f_\text{ov}$ is a user-adjusted dimensionless parameter set to 0.01.
For small convective cores, 
MESA allows the implementation of convective overshoot through a geometrical cut-off \citep{Paxton2011} that assures that overshoot is always limited to a fraction of the extension of the convective region. This is done by replacing $H_\text{P}$ in Equation (\ref{Eq.Convective_Overshoot_Parameter}) by
\begin{equation}
    \tilde{H}_P := H_p \cdot \min \left( 1, \frac{\Delta R_\text{cz}}{H_P \cdot \alpha_\text{ov}}\right),
    \label{Eq.Geometrical_Cutoff_of_Pressure_Scale_Height}
\end{equation}
where $\Delta R_\text{cz}$ is the thickness of the convectively-unstable region and $\alpha_\mathrm{ov}$ is by default set to the mixing length parameter.
Microscopic diffusion was not included. 

The stellar models for all evolutionary tracks from the Hayashi line to the bottom of the red-giant branch were computed using MESA, but only the models from the subgiant phase were saved. This stage of evolution corresponds to the models calculated between the Terminal Age of the Main Sequence (TAMS), defined as when $\log(\rho_c)\simeq3.0$ g.cm$^{-3}$ (corresponding approximately to the age at which the central hydrogen abundance is $X_\text{c}=10^{-4}$ in our grid), and the Terminal Age of the SubGiant phase (TASG), defined as when $\log(\rho_c)\simeq4.5$ g.cm$^{-3}$ \citep{Nsamba2018}. Although unconventional, this $\rho_\text{c}$-condition for TAMS allows us to structure the grid with a constant $\rho_\text{c}$-step, starting from models with the same $\rho_\text{c}$ value.

\begin{figure}
    \resizebox{\hsize}{!}{\includegraphics{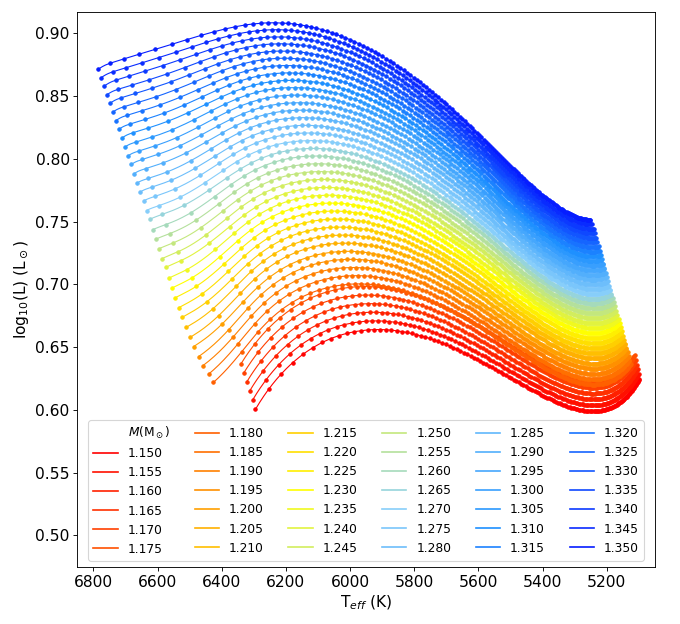}}
    \caption{HR diagram displaying every 10th model of each of the 41 tracks in the subgiant phase produced for the stellar grid described in the text.}
    \label{Fig.HR_Diagram_of_Initial_Grid}
\end{figure}

Computing a grid of stellar models capable of comprehensively describing oscillation modes in the subgiant phase is time-consuming as mixed modes vary rapidly with age. We programmed MESA to output models at a $\rho_c$-step of approximately 10 g.cm$^{-3}$, which resulted in a dense grid of 121,424 stellar models distributed over the 41 evolutionary tracks, with the number of models per evolutionary track varying between 2959 and 2963. 
As the models of our stellar grid are equally spaced in central density, the number of models in the later stages of evolution is much larger than the number of models in the earlier stages of evolution, as seen in Figure \ref{Fig.HR_Diagram_of_Initial_Grid}, 
where we plot one in every 10 models along evolutionary tracks for illustration purposes. This information is important in order to correctly interpret the results we will present in our figures and tables.

For each model generated by MESA, we used GYRE \citep{Townsend2013} to compute theoretical stellar oscillation frequencies for modes with $\ell$ = 0 and 1, by solving the equations for adiabatic stellar pulsations, outputting the frequencies within the frame window determined by
\begin{equation}
    [\nu_{\ell,n}] = \nu_\text{max} \pm 3\sigma_\nu \text{, where } \sigma_\nu = \frac{0.66 \nu_\text{max}^{0.88}}{2\sqrt{2\log(2)}}
    \label{Eq.GYRE_Frame_Window}
\end{equation}
where $\nu_\text{max}$ is the frequency of maximum power, and $\sigma_\nu$ is the width of the Gaussian envelope encompassing the distribution of pulsations, as given by \cite{Mosser2012a}. 

While all radial orders of $\ell$=0 modes determined by GYRE within the frame window defined by Equation (\ref{Eq.GYRE_Frame_Window}) were considered, only positive radial orders of $\ell=1$ modes were used in the tests.
Only non-radial modes of degree $\ell$=1 were computed since the objective is to compare the interpolation of pure acoustic modes ($\ell$=0) with the interpolation of mixed modes (of which $\ell$=1 modes are the best example, since their p-g coupling is maximal).


\subsection{Interpolation methods}
\label{Sec.Classical_Interpolation_Methods}

A complete interpolation process requires both interpolation along the evolutionary track (i.e., in the age proxy), and across the input parameters of the grid (in this case, mass).
Therefore, we considered a series of interpolation methods to explore the regions between stellar models, dividing the procedure in two steps: (i) age interpolation between models within an evolutionary track, considering a certain age proxy, and (ii) track interpolation across different evolutionary tracks (in the mass parameter), for a constant value of the previously chosen age proxy. 

For the interpolation procedure, we considered first (linear) 
and third (cubic) order methods, contemplating either polynomial or spline interpolation. Polynomial interpolation usually fits a $n$ order function to $n+1$ points, and is used to interpolate between the middle two. Spline interpolation considers a set of $n$ order functions, each defined between two data points to build a system of equations, constrained so that the values of the function and of its first and second derivatives match as one approaches each data point from the left and right. 
These requirements still leave the system under-constrained, which requires some reasonable assumptions to be made about the values beyond the range of data being fitted, usually an assumption about the 2nd derivative at the two end points. In both cases, the interpolated functions pass through all the existing points of the initial stellar grid. For example, consider an evolutionary track of a given mass $M$ and a mode of a given radial order $n$ (or similarly, consider a fixed value for a given age proxy $\tau$ and a mode of a given radial order $n$). Between every two models ($x_1$, $x_2$) on that evolutionary track (or at a fixed age proxy), we can linearly interpolate the frequencies to any point $x$ within that interval by computing
\begin{equation}
    f(x)=\frac{(x_2-x)y_1+(x-x_1)y_2}{x_2-x_1}
\end{equation}
where $y_1$ and $y_2$ are the mode frequencies of radial order $n$ in the models $x_1$ and $x_2$, respectively.

In the case of cubic splines, interpolation is accomplished using the function
\begin{equation}
    f(x)=
    \begin{cases}
      a_1x^3+b_1x^2+c_1x+d_1,\ \text{if}\ x\in[x_1,x_2]\\
      a_2x^3+b_2x^2+c_2x+d_2,\ \text{if}\ x\in]x_2,x_3]\\
      ...\\
      a_kx^3+b_kx^2+c_kx+d_k,\ \text{if}\ x\in]x_k,x_{k+1}]
    \end{cases}
    \label{Eq.Cubic_Splines_System_of_Equations}
\end{equation}
where each of the 4 coefficients ($a_k$, $b_k$, $c_k$, $d_k$), that define a segment $]x_k,x_{k+1}]$ of the resulting function, is obtained by solving the system of equations
\begin{equation}
    \begin{cases}
      a_kx_k^3+b_kx_k^2+c_kx_k+d_k = y_k\\
      a_kx_{k+1}^3+b_kx_{k+1}^2+c_kx_{k+1}+d_k = y_{k+1}\\
      \frac{\text{d}}{\text{d}x}f_{k-1}(x_k) = \frac{\text{d}}{\text{d}x}f_k(x_k)\\
      \frac{\text{d}^2}{\text{d}x^2}f_{k-1}(x_k) = \frac{\text{d}^2}{\text{d}x^2}f_k(x_k)\\
      \text{+ boundary conditions.}
    \end{cases}
\end{equation}
Note that each polynomial from Equation (\ref{Eq.Cubic_Splines_System_of_Equations}) is valid within the respective intervals only. For interpolation in Python, we used the library package \texttt{scipy.interpolate.interp1d()}, while considering its default conditions. This includes boundary conditions established by a ‘not-a-knot’ condition, i.e. the first two polynomials’ third derivatives are equal in the point where they touch each other. 

For each choice of the interpolation method, we can consider different age proxies for interpolation. Here we shall consider three options namely, physical age, $\tau$, scaled age, $\tau_\text{scaled}$, and central density, $\rho_c$, with 
\begin{equation}
    \tau_\text{scaled} = \frac{\tau-\tau_\text{TAMS}}{\tau_\text{TASG}-\tau_\text{TAMS}}.
    \label{Eq.Tau_Scaled_Definition}
\end{equation}

For each model $i$, the errors associated with each interpolation method and chosen age proxy are given by
\begin{equation}
    \delta\nu_{i,\ell,n,\text{norm}}=(\nu_{\ell,n,\text{norm},\text{theoretical}}-\nu_{\ell,n,\text{norm},\text{interpolation}})_i,
    \label{Eq.Normalized_Interpolation_Errors}
\end{equation}
where 'theoretical' and 'interpolation' stand for the value outputted by GYRE and by the interpolation algorithm considered, respectively. Dimensional versions of these errors are computed using the equation
\begin{equation}
    \delta\nu_{i,\ell,n}=\delta\nu_{i,\ell,n,\text{norm}}\times\sqrt{GM/R^3}.
    \label{Eq.Renormalized_Interpolation_Errors}
\end{equation}

\section{Characterization of the interpolation errors}
\label{Sec.Characterization_of_the_Interpolation_Errors}

To characterize the interpolation errors, we define two metrics, namely: 1) the maximum error for a given mode degree $\ell$ ($m_\ell$), i.e. the maximum absolute interpolation error observed for all modes in all interpolated models on the evolutionary track, and 2) the average offset of the mode degree $\ell$ ($\mu_\ell$), calculated as a weighted mean of all absolute interpolated errors, $|\delta\nu_{i,\ell,n}|$, for all modes on all interpolated models on the evolutionary track. Specifically,
\begin{equation}
    \mu_\ell = \frac{1}{\sum_n \sum_i w_{i,n}} \sum_n \sum_i |\delta\nu_{i,\ell,n}| w_{i,n}
    \label{Eq.Average_offset_definition}
\end{equation}
where 
\begin{equation}
    w_{i,n} = (\tau_{\text{scaled},i+1}-\tau_{\text{scaled},i-1})/2,
\end{equation}
where the sum in $n$ is over all the radial orders considered for an evolutionary track (cf. Section \ref{Sec.Input_Physics_of_the_Stellar_Grid}), while the sum in $i$ is over all the models for a given radial order and evolutionary track. 
The weight $w_{i,n}$ ensures that the same weight is given to similar intervals of $\tau_\text{scaled}$. This is to avoid that our metric is biased towards the evolved subgiants due to our grid having models similarly distributed in $\rho_\text{c}$, rather than in $\tau_\text{scaled}$.

In the sections that follow, we will drop the index $\ell$ in the quantities $m_\ell$ and $\mu_\ell$ and the index $i$ in $\delta\nu_{i,\ell,n}$.

\subsection{Interpolation along evolutionary tracks}

\begin{figure*}
    \centering
    \includegraphics[width=17cm]{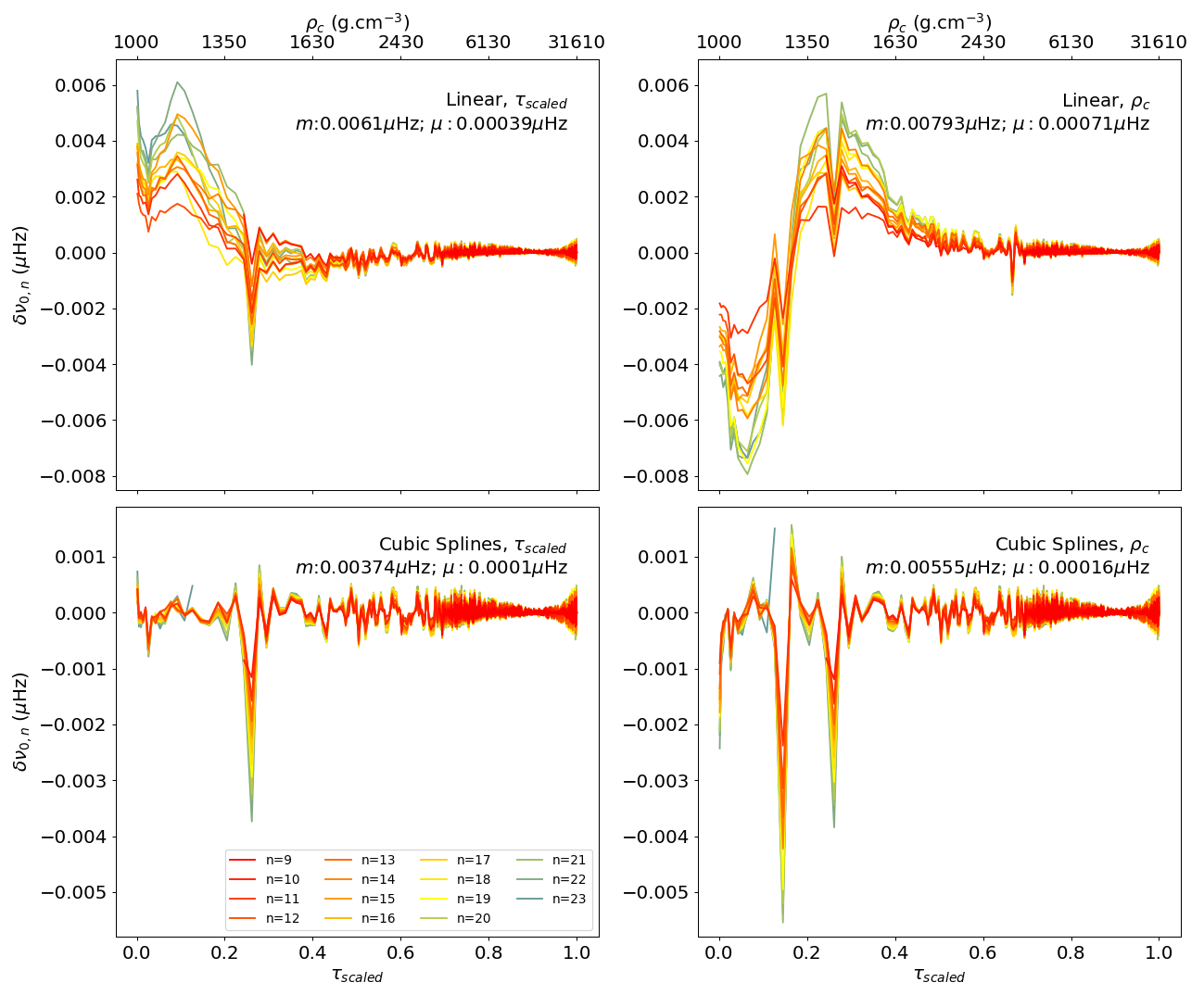}
    \caption{Interpolation errors presented as a function of $\tau_\text{scaled}$ (bottom x-axis) and $\rho_\text{c}$ (top x-axis), that resulted from interpolation along the evolutionary track with $M$=1.245 M$_\odot$, for all $\ell$=0 frequency modes, with radial orders $n\in[9,23]$, indicated by different colours. Each subplot presents the results for a different interpolation method, considering different methods - linear (top panels) or cubic splines (bottom panels) - and different age proxies - $\tau_\text{scaled}$ (left) and $\rho_\text{c}$ (right).}
    \label{Fig.Errors_for_Interpolation_Along_EvTracks_ell0}
\end{figure*}

\begin{figure*}
    \centering
    \includegraphics[width=17cm]{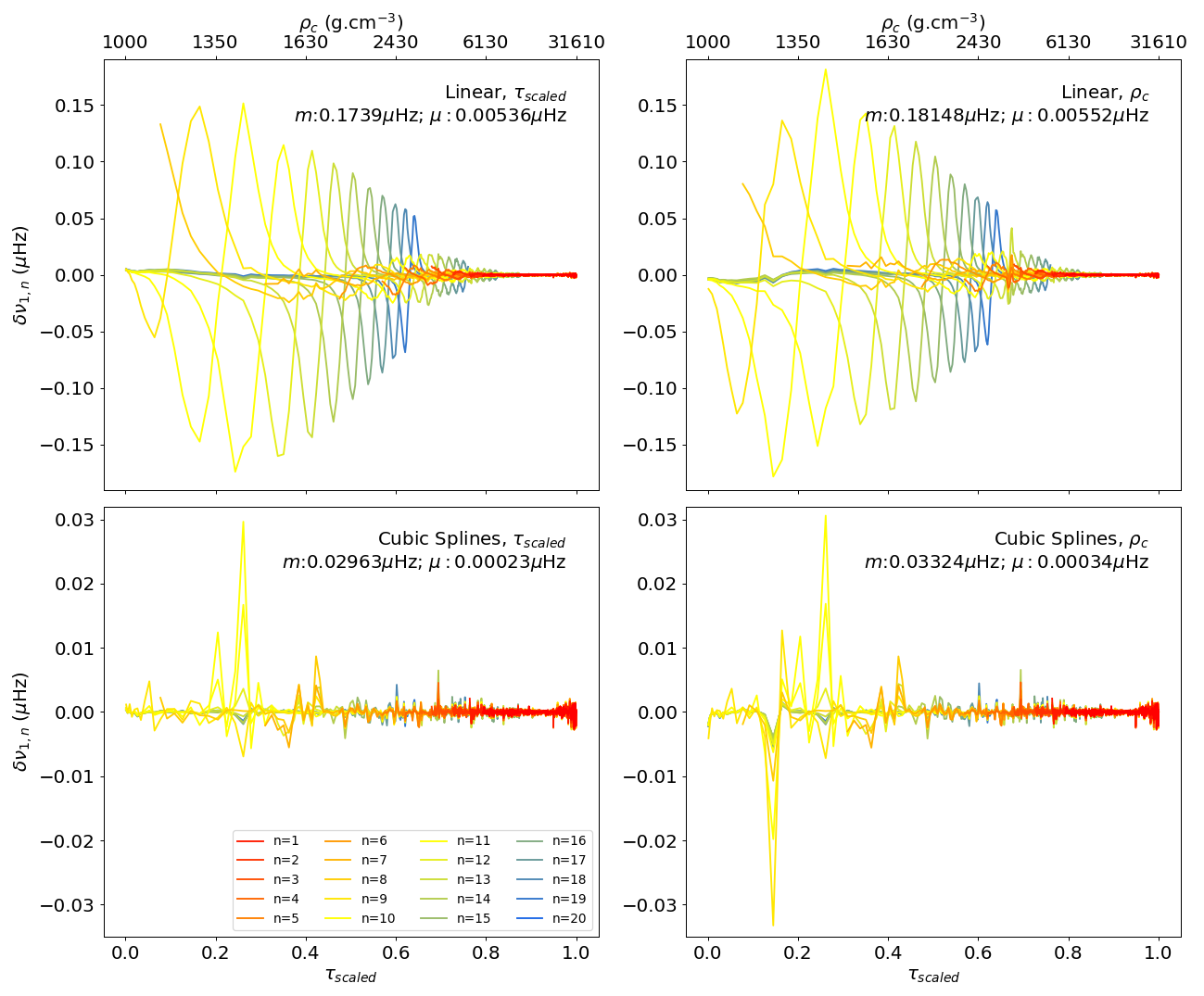}
    \caption{Interpolation errors presented as a function of $\tau_\text{scaled}$ (bottom x-axis) and $\rho_\text{c}$ (top x-axis), that resulted from interpolation along the evolutionary track with $M$=1.245 M$_\odot$, for all $\ell$=1 frequency modes, with radial orders $n\in[1,20]$, indicated by different colours. Each subplot presents the results for a different interpolation method, namely linear (top panels) and cubic splines (bottom panels), and different age proxies - $\tau_\text{scaled}$ (left) and $\rho_\text{c}$ (right).}
    \label{Fig.Errors_for_Interpolation_Along_EvTracks_ell1}
\end{figure*}

We start by performing a control study of the errors on the frequencies that result from applying interpolation along an evolutionary track. Accordingly, we consider all evolution frequency curves (each defined by a different fixed radial order) determined for the mass track with $M$=1.245 M$_\odot$, and contemplate two possible interpolation methods - linear interpolation and cubic splines, - and three possible age proxies - $\tau$, $\tau_\text{scaled}$ and $\rho_\text{c}$.
The results are displayed in Figure \ref{Fig.Errors_for_Interpolation_Along_EvTracks_ell0} and Table \ref{Tab.Maxima_and_Average_Errors_for_Interpolation_Along_EvTracks_ell0} for $\ell=0$, and in Figure \ref{Fig.Errors_for_Interpolation_Along_EvTracks_ell1} and Table \ref{Tab.Maxima_and_Average_Errors_for_Interpolation_Along_EvTracks_ell1} for $\ell=1$, according to their maximum errors, $m$, and average offsets, $\mu$, defined above.

\begin{table}[]
    \centering
    \caption{Average offsets ($\mu$) and maximum errors ($m$) from interpolation along the evolutionary track with $M$=1.245 M$_\odot$, for all $\ell=0$ mode frequencies, considering the indicated interpolation algorithms. Notice that interpolation errors are the same for $\tau$ and $\tau_\text{scaled}$.}
    \label{Tab.Maxima_and_Average_Errors_for_Interpolation_Along_EvTracks_ell0}
    \begin{tabular}{cccc}
         \hline \noalign{\smallskip}
         \textcolor{white}{-} & Errors & Linear & Cubic Splines \\
         \noalign{\smallskip} \hline \noalign{\smallskip}
         \multirow{2}{4em}{$\tau$, $\tau_\text{scaled}$} & $\mu$ ($\mu$Hz) & $3.9\times10^{-4}$ & $9.8\times10^{-5}$ \\
         & $m$ ($\mu$Hz)  & $6.1\times10^{-3}$ & $3.7\times10^{-3}$ \\ 
         \noalign{\smallskip} \hline \noalign{\smallskip}
         \multirow{2}{4em}{$\rho_c$} & $\mu$ ($\mu$Hz) & $7.1\times10^{-4}$ & $1.6\times10^{-4}$ \\
         & $m$ ($\mu$Hz) & $7.9\times10^{-3}$ & $5.6\times10^{-3}$ \\
         \noalign{\smallskip} \hline
    \end{tabular}
\end{table}

The results of interpolation along evolutionary tracks considering $\tau$ as the age proxy are exactly the same as when considering $\tau_\text{scaled}$ as the age proxy, since both age proxies are related by a proportionality constant (cf. Equation \ref{Eq.Tau_Scaled_Definition}), that does not vary within a fixed evolutionary track.

For radial modes, the average offset of the interpolation errors is about 3 orders of magnitude smaller than the accuracy limit we set ($\sim$0.1 $\mu$Hz). This is to be expected, since the frequencies of radial modes evolve smoothly and, thus, do not produce significant interpolation errors. We observe only a slight improvement when increasing the order of the interpolation (from linear to cubic splines) and do not observe a significant dependence on the adopted age proxy.

\begin{table}[]
    \centering
    \caption{Average offsets ($\mu$) and maximum errors ($m$) from interpolation along the evolutionary track with $M$=1.245 M$_\odot$, for all $\ell=1$ mode frequencies, considering the indicated interpolation algorithms. Notice that interpolation errors are the same for $\tau$ and $\tau_\text{scaled}$.}
    \label{Tab.Maxima_and_Average_Errors_for_Interpolation_Along_EvTracks_ell1}
    \begin{tabular}{cccc}
         \hline \noalign{\smallskip}
         \textcolor{white}{-} & Errors & Linear & Cubic Splines \\
         \noalign{\smallskip} \hline \noalign{\smallskip}
         \multirow{2}{4em}{$\tau$, $\tau_\text{scaled}$} & $\mu$ ($\mu$Hz) & $5.4\times10^{-3}$ & $2.3\times10^{-4}$ \\
         & $m$ ($\mu$Hz) & $1.7\times10^{-1}$ & $3.0\times10^{-2}$ \\
         \noalign{\smallskip} \hline \noalign{\smallskip}
         \multirow{2}{4em}{$\rho_c$} & $\mu$ ($\mu$Hz) & $5.5\times10^{-3}$ & $3.4\times10^{-4}$ \\
         & $m$ ($\mu$Hz) & $1.8\times10^{-1}$ & $3.3\times10^{-2}$ \\
         \noalign{\smallskip} \hline
    \end{tabular}
\end{table}

For the $\ell=1$ modes, we again observe that the interpolation errors are essentially independent of the age proxy considered. 
However, the results for the dipole modes show a significant dependence on the interpolation method, with the average offset resulting from linear interpolation being $\sim$19.8 times larger than the average offset obtained using cubic splines. 
This shows that the cubic splines interpolation method is preferred for interpolation along evolutionary tracks for dipole modes. Moreover, the results show that the maximum error when interpolating with cubic splines along the evolutionary track is about an order of magnitude smaller than the typical observational errors ($\sim$0.1 $\mu$Hz), leaving room for additional errors that will result from the interpolation across different evolutionary tracks, and that we will address in Sections \ref{Sec.The_Interpolation_Method_with_tau_age_proxies}-\ref{Sec.Interpolation_Method_with_different_order_resolutions}.

We also emphasize that the results for the mass-track considered for these tests, with $M$=1.245 M$_\odot$, are of similar magnitude to those obtained for other mass-tracks, making this evolutionary track a good representative for interpolation along tracks.


\subsection{Full Interpolation with $\tau$ and $\tau_\text{scaled}$ as age proxies}
\label{Sec.The_Interpolation_Method_with_tau_age_proxies}

In this section, we consider the errors resulting from the application of the full interpolation algorithm employed in the computation of the frequencies of models that do not lie on a particular evolutionary track.
To that end, we consider a subset of our grid composed only of the models whose third decimal place of the associated mass value is 0 (e.g. 1.240) in order to retrieve through interpolation the models of the initial complete grid (namely those whose third decimal place of the associated mass value is 5, like 1.245).

\begin{figure}
    \resizebox{0.97\hsize}{!}{\includegraphics{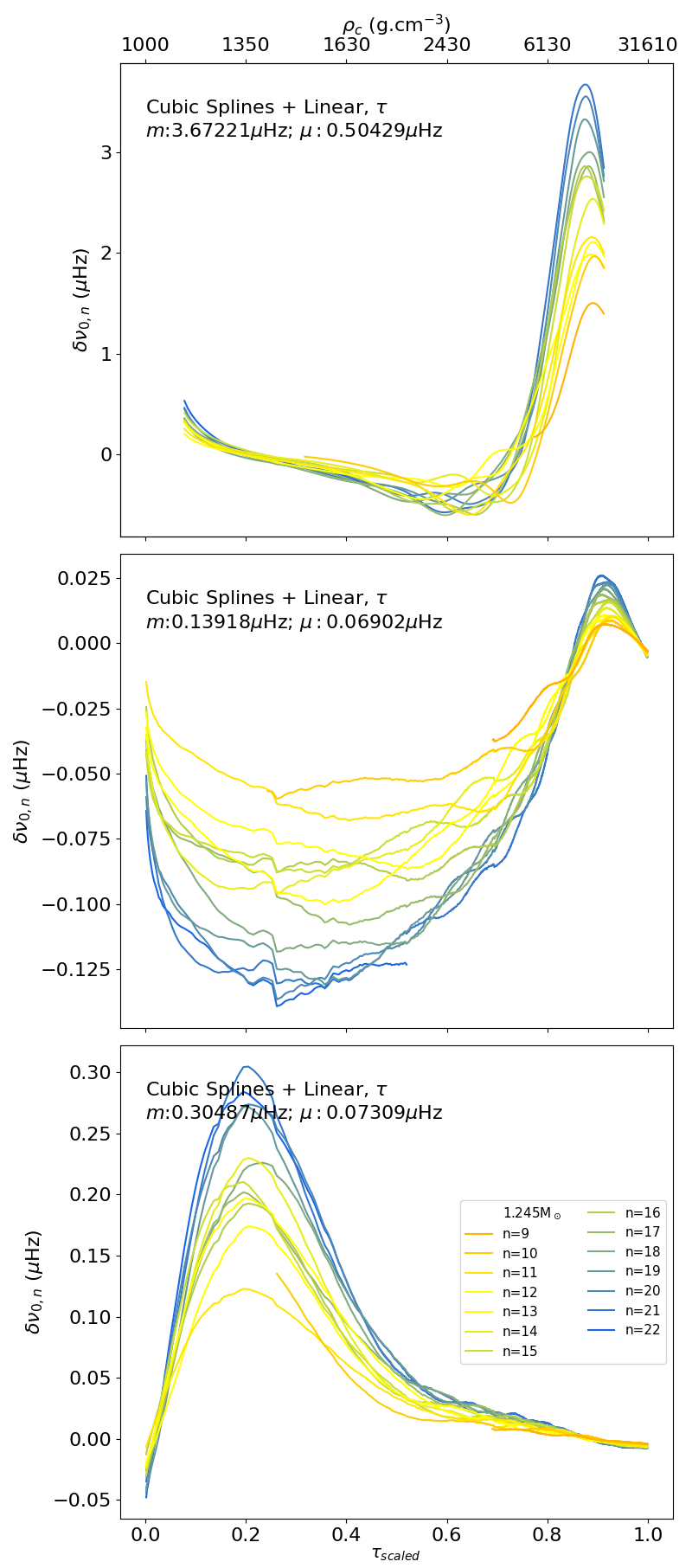}}
    \caption{Interpolation errors presented as a function of $\tau_\text{scaled}$ (bottom x-axis) and $\rho_\text{c}$ (top x-axis), that resulted from the full interpolation algorithm (cubic splines interpolation along tracks, and linear interpolation across tracks) for all $\ell$=0 modes associated with the mass track $M$=1.245 M$_\odot$, when considering $\tau$ (top), $\tau_\text{scaled}$ (middle) and $\rho_\text{c}$ (bottom) as the age proxies for interpolation.}
    \label{Fig.Linear_Interpolation_Across(l=0)}
\end{figure}

\begin{figure}
    \resizebox{0.97\hsize}{!}{\includegraphics{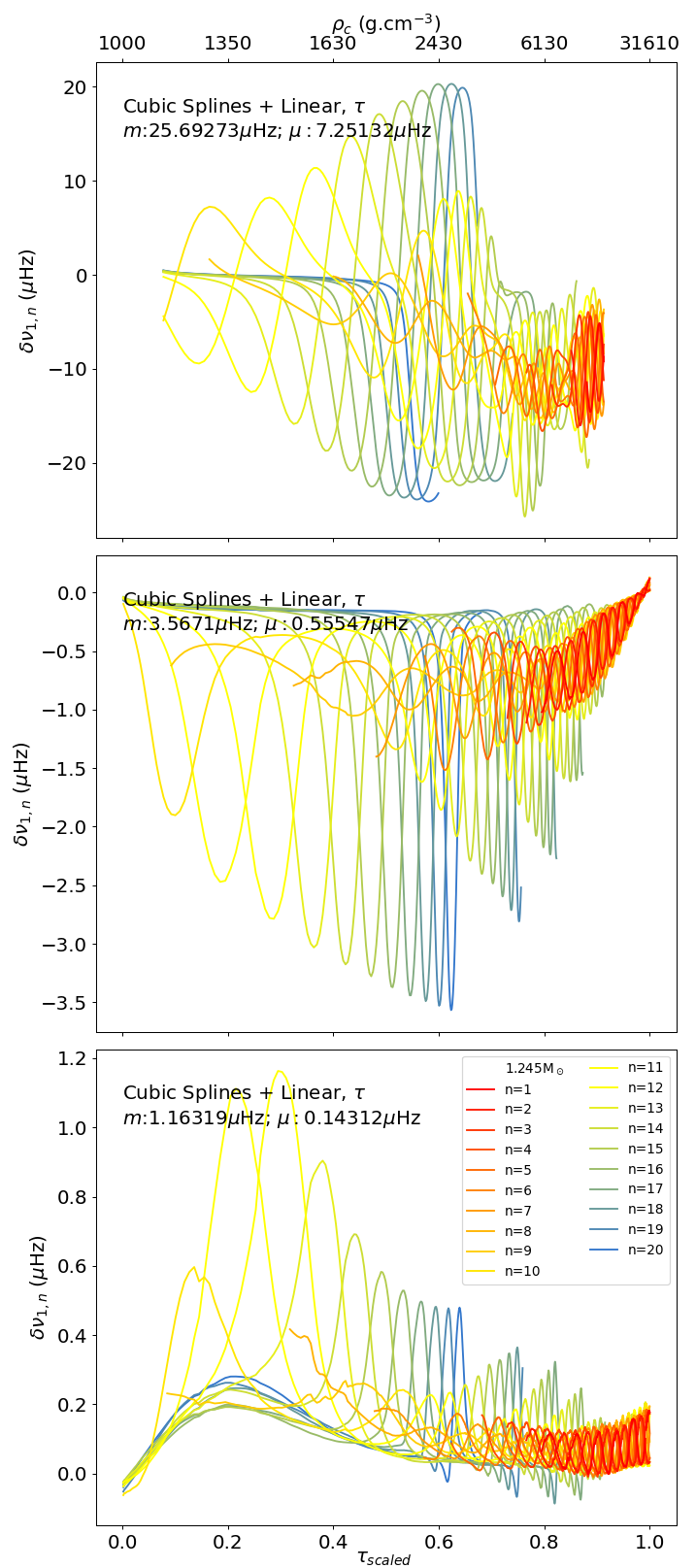}}
    \caption{Interpolation errors presented as a function of $\tau_\text{scaled}$ (bottom x-axis) and $\rho_\text{c}$ (top x-axis), that resulted from the full interpolation algorithm (cubic splines interpolation along tracks, and linear interpolation across tracks) for all $\ell$=1 modes associated with the track with $M$=1.245 M$_\odot$, when considering $\tau$ (top), $\tau_\text{scaled}$ (middle) and $\rho_\text{c}$ (bottom) as the age proxies for interpolation.}
    \label{Fig.Linear_Interpolation_Across(l=1)}
\end{figure}

\begin{figure}
    \resizebox{0.97\hsize}{!}{\includegraphics{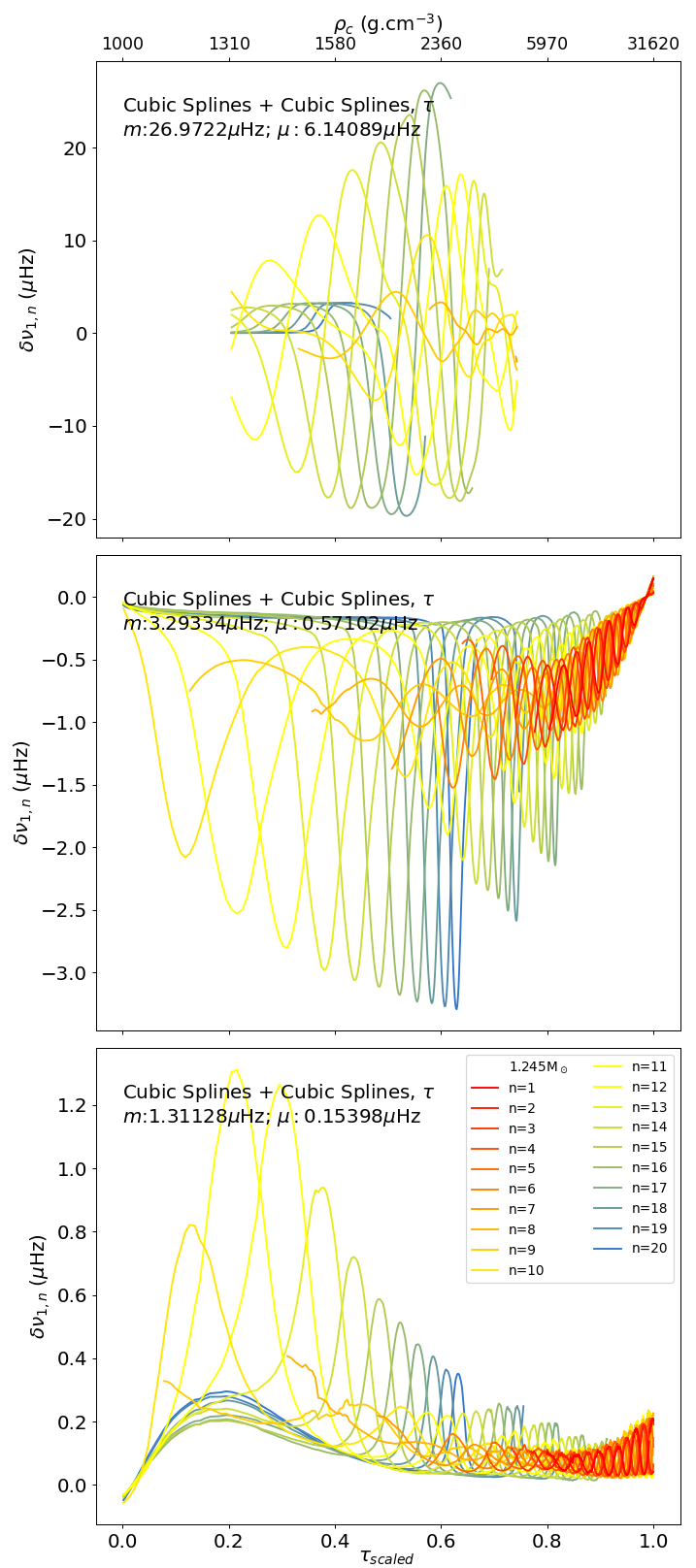}}
    \caption{Interpolation errors presented as a function of $\tau_\text{scaled}$ (bottom x-axis) and $\rho_\text{c}$ (top x-axis), that resulted from the full interpolation algorithm (cubic splines interpolation both along and across tracks) for all $\ell$=1 modes associated with the track with $M$=1.245 M$_\odot$, when considering $\tau$ (top), $\tau_\text{scaled}$ (middle) and $\rho_\text{c}$ (bottom) as the age proxies for interpolation.}
    \label{Fig.Cubic_Interpolation_Across(l=1)}
\end{figure}

As before, the initial tests are performed for models with $M$=1.245 M$_\odot$, whose frequencies are computed through interpolation, and subsequently compared with the true frequencies computed in the initial grid with GYRE. These tests will then be extended to the full grid in Section \ref{Sec.Interpolation_Results_of_Full_Grid} and in Appendix \ref{AppA:Errors_of_Interpolated_Mass_Tracks_of_Grid}.
The interpolation is made using the corresponding adjacent evolutionary tracks from the subset grid (2 for linear interpolation across tracks and 4 for the cubic splines interpolation across tracks).
The results considering cubic splines along an evolutionary track and linear interpolation across evolutionary tracks are shown in Figure \ref{Fig.Linear_Interpolation_Across(l=0)} and Table \ref{Tab.Maxima_and_Average_Errors_for_Interpolation_Across_EvTracks_l0} for $\ell$=0, and in Figure \ref{Fig.Linear_Interpolation_Across(l=1)} and Table \ref{Tab.Maxima_and_Average_Errors_for_Interpolation_Across_EvTracks} for $\ell$=1.
Results from interpolating $\ell$=1 modes using cubic splines both along and across evolutionary tracks are shown in Figure \ref{Fig.Cubic_Interpolation_Across(l=1)}. In all three figures, the results from considering $\tau$ as the age proxy are shown on the top panels, and the results from considering $\tau_\text{scaled}$ as the age proxy are shown in the middle panels.

\begin{figure}
    \resizebox{\hsize}{!}{\includegraphics{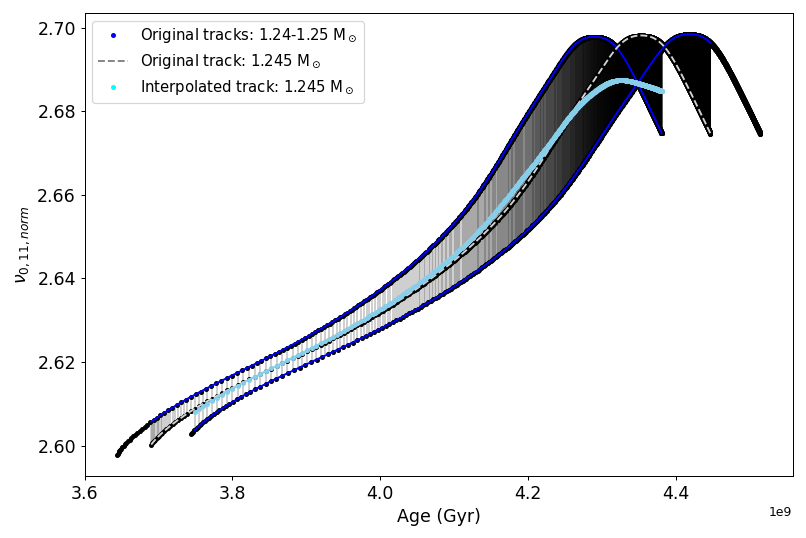}}
    \resizebox{\hsize}{!}{\includegraphics{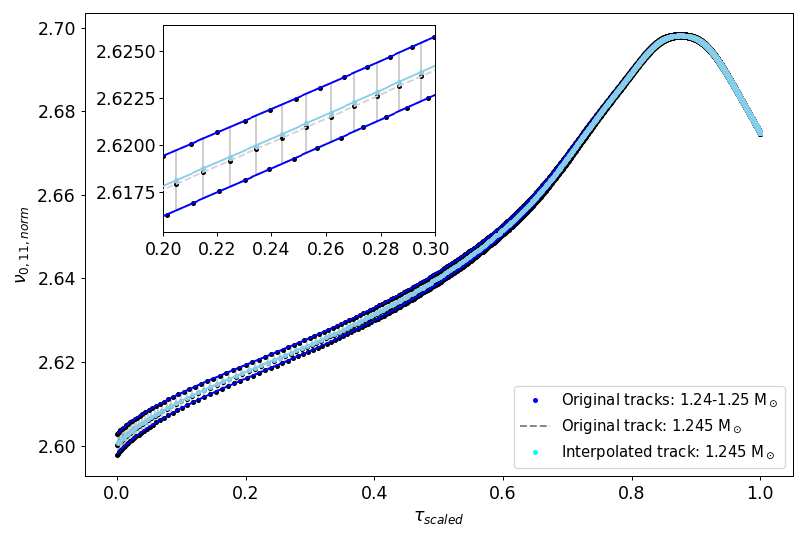}}
    \caption{Evolution of the scaled frequencies for the $\ell=0$ modes, with radial order $n$=11, that resulted from applying the interpolation algorithm to the mass-track with $M$=1.245 M $_\odot$, presented as a function of $\tau$ (top) and $\tau_\text{scaled}$ (bottom), when each is considered as the respective age proxy for interpolation. The bottom panel displays a subplot that zooms on the region of the age proxy parameter space with the highest interpolation errors.}
    \label{Fig.:Interpolation_studies_for_l0_modes}
\end{figure}

For radial modes (Figure \ref{Fig.Linear_Interpolation_Across(l=0)}), the best maximum errors and average offsets were obtained when $\tau_\text{scaled}$ is considered as the age proxy, with values of $1.4\times10^{-1}$ and $6.9\times10^{-2}$ $\mu$Hz, respectively. This shows that, even for radial modes, interpolation errors can be larger than our goal, although not significantly so.
Despite the absence of the avoided crossing phenomena, the evolutions of $\ell$=0 mode frequencies for different mass tracks are not aligned when described as a function of the age proxies considered, thus, interpolation between different evolutionary tracks will still happen between different phases of evolution of the modes, as shown in Figure \ref{Fig.:Interpolation_studies_for_l0_modes}. This is more evident for the frequency evolution in age (top panel) than for the frequency evolution in $\tau_\text{scaled}$ (bottom panel), which is consistent with the errors presented in Table \ref{Tab.Maxima_and_Average_Errors_for_Interpolation_Across_EvTracks_l0}.

\begin{figure*}
    \centering
    \includegraphics[width=17cm]{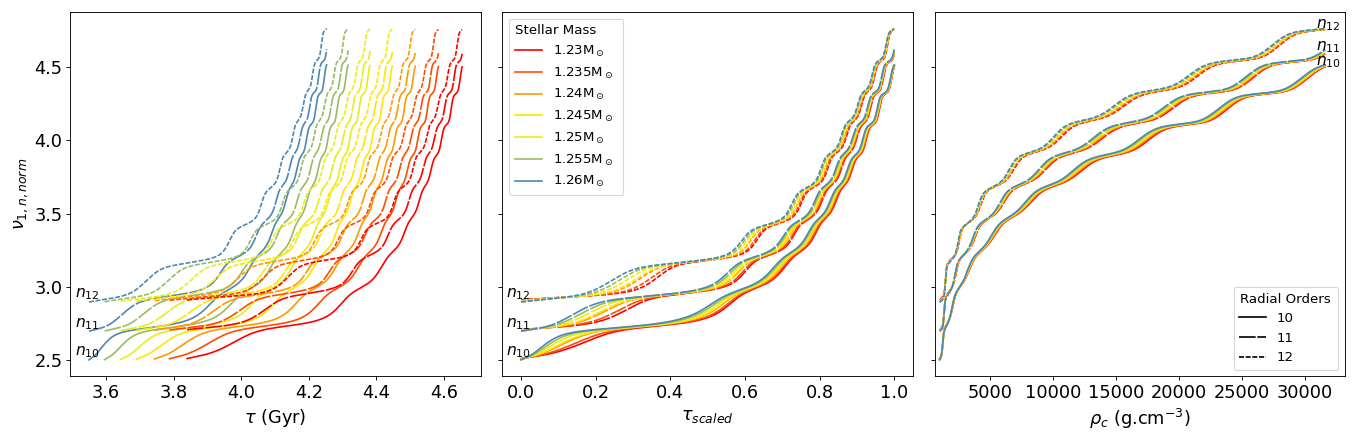}
    \caption{Evolution of the normalized frequencies of non-radial $\ell$=1 modes, with radial orders $n$=[10,11,12], for evolutionary tracks of the stellar grid with masses between 1.23 and 1.26 M$_\odot$, as a function of the various age proxies considered for interpolation: $\tau$ (left), $\tau_\text{scaled}$ (middle), and $\rho_c$ (right). Different colours are used to represent different stellar masses and different line styles show different radial orders, according to the insets in the middle and right panel, respectively.}
    \label{Fig.Phase_Comparison_of_ell1_Modes}
\end{figure*}

In the case of $\ell=1$ modes, the linear interpolation across tracks is again found to give better results when $\tau_\text{scaled}$ is taken as the age proxy (middle panel of Figure \ref{Fig.Linear_Interpolation_Across(l=1)}), this compared to using $\tau$ (top panel of Figure \ref{Fig.Linear_Interpolation_Across(l=1)}). 
However, the average offset of $5.6\times10^{-1}$ $\mu$Hz and the maximum error of $3.6\times10^0$ $\mu$Hz are already larger than our accuracy limit, with $m$ being an order of magnitude larger than that limit. 
The reason for these errors when considering $\tau$ or $\tau_\text{scaled}$ as age proxy has been alluded to in Section \ref{Sec.Impact_of_Mixed_Modes_in_Subgiants} and can be further understood from the left and middle panels of Figure \ref{Fig.Phase_Comparison_of_ell1_Modes}: because the mixed modes are not aligned when the frequency curves are described as a function of these age proxies, interpolation between different evolutionary tracks happens between different phases of evolution of the mixed modes, failing even when considering a relatively dense grid of models. 
Thus, for the cases presented, it is not possible to obtain results within the accuracy limit set by the typical observational errors, with this choice of methods and age proxies, which could potentially limit the use of stellar grids with interpolation to explore the parameter space when fitting seismic data for subgiant stars.

\begin{table}
    \centering
    \caption{Average offset ($\mu$) and maximum errors ($m$) from the full interpolation algorithm (that considers cubic splines along tracks and linear across tracks). The results are for all $\ell$=0 modes associated with the evolutionary track with $M$=1.245 M$_\odot$ (determined from interpolation using adjacent evolutionary tracks), for each of the age proxies indicated.}
    \label{Tab.Maxima_and_Average_Errors_for_Interpolation_Across_EvTracks_l0}
    \begin{tabular}{ccc}
         \hline \noalign{\smallskip}
         \textcolor{white}{-} & Errors & Linear \\
         \noalign{\smallskip} \hline \noalign{\smallskip}
         \multirow{2}{4em}{$\tau$} & $\mu$ ($\mu$Hz) & $5.0\times10^{-1}$ \\
         & $m$ ($\mu$Hz) & $3.7\times10^{0}$ \\
         \noalign{\smallskip} \hline \noalign{\smallskip}
         \multirow{2}{4em}{$\tau_\text{scaled}$} & $\mu$ ($\mu$Hz) & $6.9\times10^{-2}$ \\
         & $m$ ($\mu$Hz) & $1.4\times10^{-1}$ \\
         \noalign{\smallskip} \hline \noalign{\smallskip}
         \multirow{2}{4em}{$\rho_c$} & $\mu$ ($\mu$Hz) & $7.3\times10^{-2}$ \\
         & $m$ ($\mu$Hz) & $3.0\times10^{-1}$ \\
         \noalign{\smallskip} \hline
    \end{tabular}
\end{table}

\begin{table}[]
    \centering
    \caption{Average offset ($\mu$) and maximum errors ($m$) from the full interpolation algorithm (that considers cubic splines along tracks and the indicated regression method across tracks). The results are for all $\ell$=1 modes associated with the evolutionary track with $M$=1.245 M$_\odot$ (determined from interpolation using adjacent evolutionary tracks), for each of the age proxies indicated.}
    \label{Tab.Maxima_and_Average_Errors_for_Interpolation_Across_EvTracks}
    \begin{tabular}{cccc}
         \hline \noalign{\smallskip}
         \textcolor{white}{-} & Errors & Linear & Cubic Splines \\
         \noalign{\smallskip} \hline \noalign{\smallskip}
         \multirow{2}{4em}{$\tau$} & $\mu$ ($\mu$Hz) & $7.3\times10^{0}$ & $6.1\times10^{0}$ \\
         & $m$ ($\mu$Hz) & $2.6\times10^{1}$ & $2.7\times10^{1}$ \\
         \noalign{\smallskip} \hline \noalign{\smallskip}
         \multirow{2}{4em}{$\tau_\text{scaled}$} & $\mu$ ($\mu$Hz) & $5.6\times10^{-1}$ & $5.7\times10^{-1}$ \\
         & $m$ ($\mu$Hz) & $3.6\times10^{0}$ & $3.3\times10^{0}$ \\
         \noalign{\smallskip} \hline \noalign{\smallskip}
         \multirow{2}{4em}{$\rho_c$} & $\mu$ ($\mu$Hz) & $1.4\times10^{-1}$ & $1.5\times10^{-1}$ \\
         & $m$ ($\mu$Hz) & $1.2\times10^{0}$ & $1.3\times10^{0}$ \\
         \noalign{\smallskip} \hline
    \end{tabular}
\end{table}

\subsection{The central density $\rho_\text{c}$ as an alternative age proxy}
\label{Sec.The_Interpolation_Method_with_rhoc_age_proxies}

As discussed in Section \ref{Sec.Impact_of_Mixed_Modes_in_Subgiants}, the avoided crossings of evolutionary tracks with different masses appear more aligned when displayed as a function of the central density, $\rho_\text{c}$ (right plot of Figure \ref{Fig.Phase_Comparison_of_ell1_Modes}).
With the avoided crossings showing an increased alignment when described as a function of $\rho_\text{c}$, interpolation can happen between models whose mode frequencies are in a more similar phase of evolution, which is expected to lead to more accurate results.

With the above in mind, we now analyse the results from applying the full interpolation algorithm to interpolate the frequencies of models of the evolutionary track of $M$=1.245 M$_\odot$, using $\rho_\text{c}$ as the age proxy.
These results are shown in the bottom panels of Figures \ref{Fig.Linear_Interpolation_Across(l=0)}, \ref{Fig.Linear_Interpolation_Across(l=1)} and \ref{Fig.Cubic_Interpolation_Across(l=1)}, and summarized in Tables \ref{Tab.Maxima_and_Average_Errors_for_Interpolation_Across_EvTracks_l0} and \ref{Tab.Maxima_and_Average_Errors_for_Interpolation_Across_EvTracks}, for mode frequencies with $\ell$=0 and 1, respectively.

For the radial modes ($\ell$=0) (bottom panel of Figure \ref{Fig.Linear_Interpolation_Across(l=0)}), interpolation errors using $\rho_\text{c}$ as the age proxy can be as high as $3.0\times10^{-1}$ $\mu$Hz, only decreasing below $5.0\times10^{-2}$ $\mu$Hz in later stages of evolution. 
In fact, this maximum error is close but slightly larger than the one obtained considering $\tau_\text{scaled}$ as the age proxy ($m=1.4\times10^{-1}$ $\mu$Hz) and than the reference value of 0.1 $\mu$Hz. This is not surprising since radial modes do not have avoided crossings, thus no particular improvement was to be expected by choosing $\rho_\text{c}$ over $\tau_\text{scaled}$ as the age proxy.
Although some of the radial orders produce interpolation errors above the established accuracy limit, these are limited to the models of the early subgiant phase with $\tau_\text{scaled}<$ 0.4. Despite these large maximum absolute errors in some models, the average offset is $7.3\times10^{-2}$ $\mu$Hz, well within the proposed accuracy limit. 

For $\ell=1$ modes, errors from interpolation considering $\rho_\text{c}$ as the age proxy can be as high as $1.2\times10^0$ and $1.3\times10^0$ $\mu$Hz, depending on the order of the interpolation method considered (Figures \ref{Fig.Linear_Interpolation_Across(l=1)} and \ref{Fig.Cubic_Interpolation_Across(l=1)}, respectively), only decreasing to values below $4.0\times10^{-1}$ $\mu$Hz in the later stages of the subgiant evolution.
Although their maxima are quite large (about an order of magnitude above the accuracy limit we set as reference), the average offset is improved by a factor of $\sim$3.9 over the average offset of interpolation with $\tau_\text{scaled}$ as the age proxy, yielding results similar to the expected accuracy limits - $1.4\times10^{-1}$ and $1.5\times10^{-1}$ $\mu$Hz for linear and cubic splines interpolation, respectively.
We verified that this improvement in the interpolation errors is not a consequence of the grid being equally spaced in $\rho_\text{c}$, rather than $\tau_\text{scaled}$, (notice that the errors from interpolation along the tracks are always small, regardless of the age proxy being adopted when compared with the final errors (cf. Table \ref{Tab.Maxima_and_Average_Errors_for_Interpolation_Along_EvTracks_ell1})).


\subsection{Results for the full grid}
\label{Sec.Interpolation_Results_of_Full_Grid}

Next, we apply the analysis to the full grid and discuss the complete set of results. To this end, we use the mode frequencies determined as part of the subset (whose third decimal place of the associated mass value is 0) to obtain the frequencies for the missing mass-tracks that also appear in the initial grid (i.e. those whose third decimal place of the associated mass value is 5).
Results are presented in the tables in Appendix \ref{AppA:Errors_of_Interpolated_Mass_Tracks_of_Grid}, and summarized in Figure \ref{Fig.Interpolated_Errors_for_the_Full_Grid}. Interpolation was not performed for tracks with masses 1.155 and 1.345 M$_\odot$, given that they are near the limit where there are not enough models on one of the sides to allow cubic spline interpolation.

The results from Figure \ref{Fig.Interpolated_Errors_for_the_Full_Grid} show that using $\rho_\text{c}$ as the age proxy significantly improves the accuracy of the interpolation when compared to $\tau_\text{scaled}$.
For $M$>1.20 M$_\odot$, both the average offset and maximum error are smaller when $\rho_\text{c}$ is used, instead of $\tau_\text{scaled}$, as the age proxy. For the smaller masses, there are a few cases when the interpolation errors obtained with the two age proxies are comparable, but only for one mass-track both the average offset and the maximum error are larger when $\rho_\text{c}$ is chosen.
Moreover, we see that, with the exception of the mass track $M$=1.175 M$_\odot$ (brown dashed line in Figure \ref{Fig.Interpolated_Errors_for_the_Full_Grid}), the average offset when taking $\rho_\text{c}$ as the age proxy is always approximately within twice of the reference value.
It also becomes evident that for the more massive evolutionary tracks the average offset when considering $\tau_\text{scaled}$ as the age proxy is similar to the maximum error found when $\rho_\text{c}$ is chosen. Thus, when considering the full grid, the adoption of $\rho_\text{c}$ in place of $\tau_\text{scaled}$ as the age proxy seems well justified. 
Finally, Figure \ref{Fig.Interpolated_Errors_for_the_Full_Grid} confirms that the evolutionary track of $M$=1.245 M$_\odot$ (brown dotted line), considered in the previous sections, has typical interpolation errors, justifying its use in the more detailed analysis presented, and it also shows that there is an evolutionary track ($M$=1.175 M$_\odot$) for which interpolation performs significantly worse. We shall address the origin of the anomalous large interpolation errors found for the latter model in Section \ref{Sec.Influence_of_Core_Properties_in_Interpolation}.


\subsection{Impact of the order of interpolation}
\label{Sec.Interpolation_Method_with_different_order_resolutions}

Having discussed the results of interpolation considering different age proxies, we now compare the performance of the different orders of the interpolation when interpolating across evolutionary tracks, for a fixed value of the age proxy.
We compare the results of applying linear interpolation (Figure \ref{Fig.Linear_Interpolation_Across(l=1)}) and cubic splines interpolation (Figure \ref{Fig.Cubic_Interpolation_Across(l=1)}) to the $\ell$=1 modes of models in the mass track with $M$=1.245 M$_\odot$. These results (summarized in Table \ref{Tab.Maxima_and_Average_Errors_for_Interpolation_Across_EvTracks}) show that the average offset and the maximum error hardly change when changing the order of the interpolation. This is true both when adopting $\tau_\text{scaled}$ and $\rho_\text{c}$ as age proxies.

Considering that the full grid does not change these conclusions in any significant way, taking $\rho_\text{c}$ as the age proxy, we found eight interpolated mass tracks for which linear interpolation performed better (i.e. resulting in smaller $\mu$ and $m$ values than when employing the cubic splines), nine cases where cubic splines interpolation performed better, and one where the results were mixed (i.e. for $M$=1.225 M$_\odot$, the smallest $\mu$ and $m$ values were not found using the same method). These results seem to indicate no clear preference for one of the interpolation methods when performing interpolation across evolutionary tracks.
Moreover, inspection of the tables in Appendix \ref{AppA:Errors_of_Interpolated_Mass_Tracks_of_Grid} shows that the average offsets computed with linear and cubic splines interpolation on each individual track are almost always of the same order of magnitude. 
The most significant differences are found for $M$ = 1.165 and 1.185 M$_\odot$, for which $\mu$ from linear interpolation can be $\sim$6.18 and $\sim$3.19 times smaller than $\mu$ from cubic splines, respectively.
These two cases, together with the mass-track with $M$=1.175 M$_\odot$, which presents the highest interpolation errors of this grid (see Figure \ref{Fig.Interpolated_Errors_for_the_Full_Grid}), span the region of the mass parameter space where the core properties of the modelled stars change and it shall be discuss in detail in Section \ref{Sec.Influence_of_Core_Properties_in_Interpolation}.

Given the results presented in the previous subsections, we favour the interpolation option that considers the simplest approach and least computational expensive: cubic splines to interpolate along tracks, linear interpolation to interpolate across tracks, and $\rho_\text{c}$ as the age proxy.


\subsection{Testing the grid resolution}

We now recall that every evolutionary track in the initial grid has between 2959 and 2963 models, with about half of them ($\sim$ 1482 models) being used to determine the other half through interpolation. This is a large number of models occupying the memory of the system and, therefore, in what follows we will assess the impact of the grid resolution on the interpolation errors. Considering the results from the previous sections, we divide this exercise in two parts, with two different objectives. The first concerns interpolation along evolutionary tracks, where we will verify what is the minimum number of models per track for which the results remain accurate within the accuracy limit.

With that objective in mind, we consider a number of subgrids, each with a fraction of the resolution of the initial grid, $1/k$, where each evolutionary track is built by collecting every $k^\text{th}$ model of the initial track.
In this study, we focus on the evolutionary track with $M$=1.245 M$_\odot$, and test grid resolutions between 1/1 (the initial grid) and 1/6 of the initial grid (which only contains $\sim$247 models per track for interpolation). Although we still test different interpolation methods, we establish $\rho_\text{c}$ as a definitive age proxy.
The results are summarized in Table \ref{Tab.Maxima_and_Average_Errors_for_Interpolation_Along_EvTracks_Grid_Resolution}
\footnote{In addition to cubic splines, we tested the grid resolution using interpolation with cubic polynomials, for which we expect the errors to increase as $k^4$. As we found that not to be the case, we made a sanity test on an evolutionary track with 10 times the number of models of the initial grid and we found out that our interpolation routine does produce errors that increase with the expected scaling, implying that the reason this does not happen here is because our grid is not sufficiently dense to be in such regime.}. 
 
We observe that both $\mu$ and $m$ increase by a significant factor whenever the grid resolution is reduced. For a grid resolution of 1/6 of the initial grid, both the average offset $\mu$ and the maximum error $m$ increase by more than 2 orders of magnitude. However, while the average offset $\mu$ changes from $3.4\times10^{-4}$ to $1.3\times10^{-1}$ $\mu$Hz, a value still close to the established accuracy limit, the maximum errors increase from $3.3\times10^{-2}$ to $3.4\times10^0$ $\mu$Hz, this second value being significantly out of the established accuracy limit.
In the initial grid, to keep the errors of the interpolation along evolution within values comparable to the goal, our results indicate that the minimal grid resolution recommended should be no less than 1/3 of the initial grid (i.e. a grid of stellar models containing $\sim$494 models per track for interpolation). In this case, the maximum error of $2.1\times10^{-1}$ $\mu$Hz is only slightly higher than the established accuracy limit. 

The second part of the exercise concerns the impact of the grid resolution on the interpolation across evolutionary tracks. In Section \ref{Sec.Interpolation_Results_of_Full_Grid}, we have found that the results for the current grid fail to comply with the reference accuracy in what concerns the maximum error, while performing reasonably close to the goal (within a factor of two) in what concerns the average offset, when $\rho_\text{c}$ is used as the age proxy. Depending on the quality of the data, the accuracy achieved with our reference grid  may therefore still suffice, in which case it is still interesting to understand whether a similar accuracy may be achieved with a reduced number of models. To test this, we repeat the exercise of decreasing the grid resolution for the case of full interpolation to understand the consequent impact on the interpolation errors.

Table \ref{Tab.Maxima_and_Average_Errors_for_Interpolation_Across_EvTracks_Grid_Resolution} displays the results of the full interpolation algorithm applied to different grid resolutions, considering cubic splines for interpolation along, and either linear or cubic splines for the interpolation across.
For the case we previously established as the preferred option (cubic splines along and linear across), we observe a smaller relative evolution of the interpolation errors with the resolution of the subgrid than when considering interpolation along evolution alone. 
Indeed, the average offsets remain close to the limit of accuracy, increasing only slightly until the subgrid with 1/6 of the models is considered. Moreover, the maximum errors, which are found to be out of the accuracy limit already in the initial grid, hardly change when the number of models is decreased to 1/3 of the initial value. The impact of decreasing it further, to 1/6 of the initial number of models, seems to depend on the order of the method applied for the interpolation across evolutionary tracks, with the option of cubic splines performing somewhat better when the number of models is significantly decreased.

\begin{figure*}
    \centering
    \includegraphics[width=17cm]{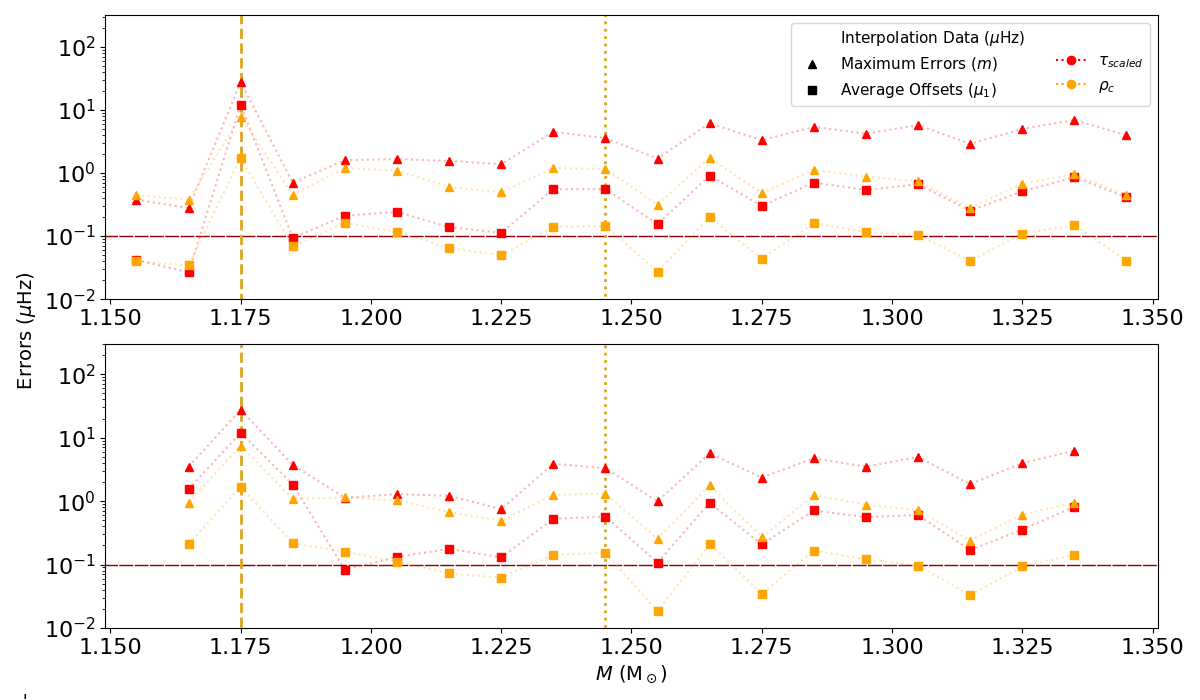}
    \caption{Average offsets ($\mu$, indicated by squares) and maximum errors ($m$, indicated by triangles) (in $\mu$Hz) from the full interpolation method (that considers cubic splines along evolutionary tracks and linear (top) or cubic splines (bottom) interpolation across evolutionary tracks, and $\tau_\text{scaled}$ (red) or $\rho_\text{c}$ (yellow) as the age proxy), for all $\ell=1$ modes obtained for the interpolated tracks of the initial grid. The vertical brown dotted line highlights the results for our reference track with $M$=1.245 M$_\odot$, while the vertical brown dashed line highlights the anomalous track with $M$=1.175 M$_\odot$.}
    \label{Fig.Interpolated_Errors_for_the_Full_Grid}
    \vspace{-1mm}
\end{figure*}

In conclusion, we find that decreasing the number of models considered in each evolutionary track to $\sim$1/3 of the initial value does not impact the accuracy of the interpolation results from our grid in any significant way. However, we emphasize that, in all cases, the maximum interpolation errors found are significantly above the goal (as seen in Figure \ref{Fig.Interpolated_Errors_for_the_Full_Grid}, and discussed in Section \ref{Sec.Interpolation_Results_of_Full_Grid}).

\begin{table}[]
    \centering
    \caption{Average offsets ($\mu$) and maximum errors ($m$) resulting from interpolation along the evolutionary track with $M$=1.245 M$_\odot$ (considering a cubic polynomial or cubic splines method), when considering $\rho_\text{c}$ as the age proxy, for all associated $\ell=1$ mode frequencies, and for the indicated grid resolutions, $1/k$.} 
    \label{Tab.Maxima_and_Average_Errors_for_Interpolation_Along_EvTracks_Grid_Resolution}
    \begin{tabular}{cccc}
         \hline \noalign{\smallskip}
         \textcolor{white}{-} & \multicolumn{3}{c}{Method for Interpolation Along}\\
         \noalign{\smallskip} \hline \noalign{\smallskip}
         Resolution & Errors & Cubic Polynomial & Cubic Splines \\
         \noalign{\smallskip} \hline \noalign{\smallskip}
         \multirow{2}{4em}{1/1} & $\mu$ ($\mu$Hz) & $7.0\times10^{-4}$ & $3.4\times10^{-4}$ \\
         & $m$ ($\mu$Hz) & $4.1\times10^{-2}$ & $3.3\times10^{-2}$ \\
         \noalign{\smallskip} \hline \noalign{\smallskip}
         \multirow{2}{4em}{1/2} & $\mu$ ($\mu$Hz) & $3.3\times10^{-3}$ & $1.0\times10^{-3}$ \\
         & $m$ ($\mu$Hz) & $1.1\times10^{-1}$ & $8.3\times10^{-2}$ \\
         \noalign{\smallskip} \hline \noalign{\smallskip}
         \multirow{2}{4em}{1/3} & $\mu$ ($\mu$Hz) & $1.3\times10^{-2}$ & $5.4\times10^{-3}$ \\
         & $m$ ($\mu$Hz) & $3.4\times10^{-1}$ & $2.1\times10^{-1}$ \\
         \noalign{\smallskip} \hline \noalign{\smallskip}
         \multirow{2}{4em}{1/4} & $\mu$ ($\mu$Hz) & $3.5\times10^{-2}$ & $1.6\times10^{-2}$ \\
         & $m$ ($\mu$Hz) & $8.3\times10^{-1}$ & $5.6\times10^{-1}$ \\
         \noalign{\smallskip} \hline \noalign{\smallskip}
         \multirow{2}{4em}{1/5} & $\mu$ ($\mu$Hz) & $7.3\times10^{-2}$ & $3.1\times10^{-2}$ \\
         & $m$ ($\mu$Hz) & $1.7\times10^0$ & $1.1\times10^0$ \\
         \noalign{\smallskip} \hline \noalign{\smallskip}
         \multirow{2}{4em}{1/6} & $\mu$ ($\mu$Hz) & $1.4\times10^{-1}$ & $1.3\times10^{-1}$ \\
         & $m$ ($\mu$Hz) & $2.8\times10^0$ & $3.4\times10^0$ \\
         \noalign{\smallskip} \hline
    \end{tabular}
\end{table}

\begin{table}[]
    \centering
    \caption{Average offsets ($\mu$) and maximum errors ($m$) from the full interpolation algorithm (that considers cubic splines along tracks, the indicated interpolation method across tracks - linear or cubic splines, - and $\rho_\text{c}$ as the age proxy) applied to all $\ell$=1 modes associated with the evolutionary track with $M$=1.245 M$_\odot$ (determined from adjacent evolutionary tracks), for different grid resolutions, $1/k$.}
    \label{Tab.Maxima_and_Average_Errors_for_Interpolation_Across_EvTracks_Grid_Resolution}
    \begin{tabular}{cccc}
         \hline \noalign{\smallskip}
         \textcolor{white}{-} & \multicolumn{3}{c}{Method for Interpolation Across}\\
         \noalign{\smallskip} \hline \noalign{\smallskip}
         Resolution & Errors & Linear & Cubic Splines \\
         \noalign{\smallskip} \hline \noalign{\smallskip}
         \multirow{2}{4em}{1/1} & $\mu$ ($\mu$Hz) & $1.4\times10^{-1}$ & $1.5\times10^{-1}$ \\
         & $m$ ($\mu$Hz) & $1.2\times10^0$ & $1.3\times10^0$ \\
         \noalign{\smallskip} \hline \noalign{\smallskip}
         \multirow{2}{4em}{1/2} & $\mu$ ($\mu$Hz) & $1.5\times10^{-1}$ & $1.6\times10^{-1}$ \\
         & $m$ ($\mu$Hz) & $1.3\times10^0$ & $1.4\times10^0$ \\
         \noalign{\smallskip} \hline \noalign{\smallskip}
         \multirow{2}{4em}{1/3} & $\mu$ ($\mu$Hz) & $1.6\times10^{-1}$ & $1.7\times10^{-1}$ \\
         & $m$ ($\mu$Hz) & $1.3\times10^0$ & $1.4\times10^0$ \\
         \noalign{\smallskip} \hline \noalign{\smallskip}
         \multirow{2}{4em}{1/4} & $\mu$ ($\mu$Hz) & $1.6\times10^{-1}$ & $1.7\times10^{-1}$ \\
         & $m$ ($\mu$Hz) & $1.5\times10^0$ & $1.4\times10^0$ \\
         \noalign{\smallskip} \hline \noalign{\smallskip}
         \multirow{2}{4em}{1/5} & $\mu$ ($\mu$Hz) & $1.7\times10^{-1}$ & $1.7\times10^{-1}$ \\
         & $m$ ($\mu$Hz) & $1.7\times10^0$ & $1.4\times10^0$ \\
         \noalign{\smallskip} \hline \noalign{\smallskip}
         \multirow{2}{4em}{1/6} & $\mu$ ($\mu$Hz) & $2.5\times10^{-1}$ & $1.8\times10^{-1}$ \\
         & $m$ ($\mu$Hz) & $3.8\times10^0$ & $1.4\times10^0$ \\
         \noalign{\smallskip} \hline
    \end{tabular}
\end{table}

\section{Influence of core properties on interpolation}
\label{Sec.Influence_of_Core_Properties_in_Interpolation}

\begin{figure}
    \resizebox{\hsize}{!}{\includegraphics{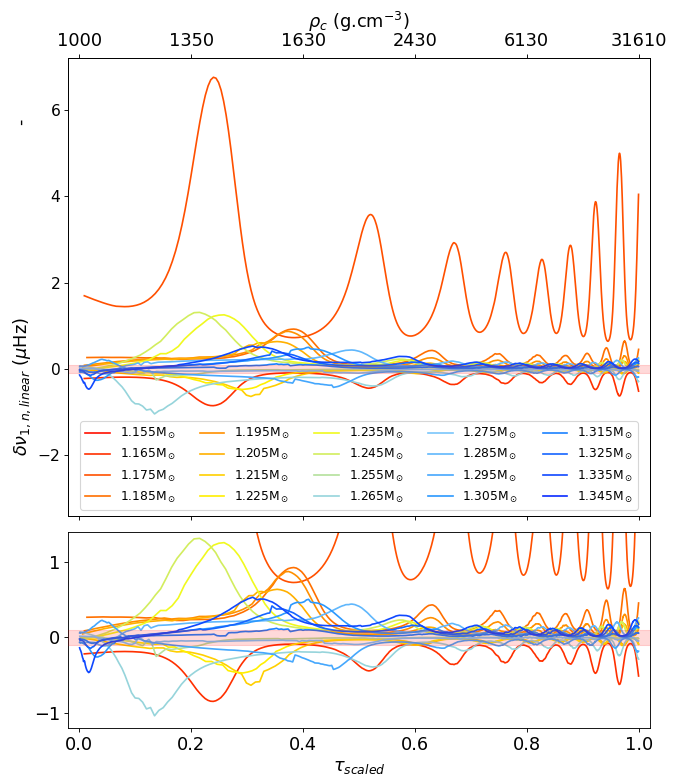}}
    \caption{Interpolation errors presented as a function of $\tau_\text{scaled}$ (bottom x-axis) and $\rho_\text{c}$ (top x-axis), that resulted from the application of the full interpolation algorithm (cubic splines along tracks and linear across tracks) across the complete grid. Results are shown for all $\nu_{1,11}$ mode frequencies, when $\rho_\text{c}$ is considered as the age proxy for interpolation. The transition mass, between radiative and convective cores in the MS, is 1.175 M$_\odot$ for this choice of $f_\text{ov}$. The bottom subplot zooms on the results for interpolation presented in the top subplot for $\delta\nu_{1,11}$ between -1.2 and 1.4 $\mu$Hz.}
    \label{Fig.Interpolation_in_rhoc_for_different_masses}
\end{figure}

\begin{figure}
    \resizebox{\hsize}{!}{\includegraphics{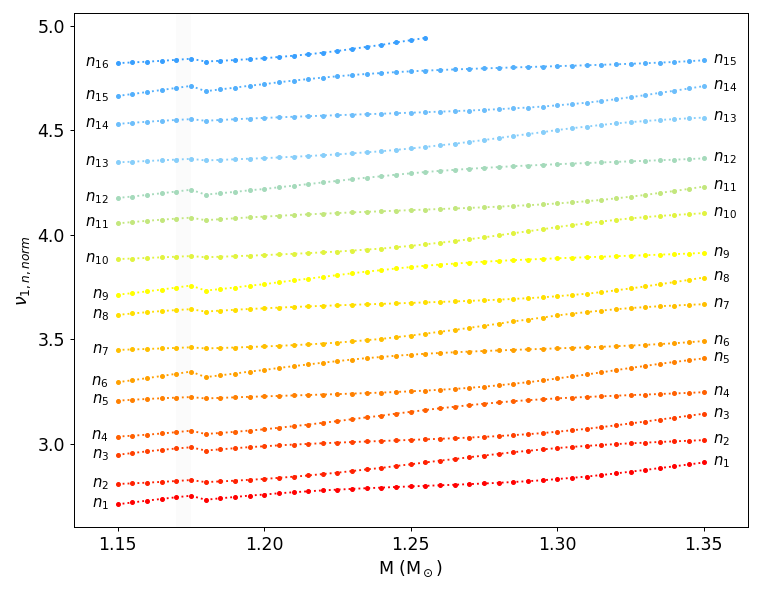}}
    \caption{Evolution of the $\ell=1$ modes with mass. Each point represents the frequency of the mode with the indicated radial order (obtained through interpolation) for the model with the given mass at the position in the evolutionary track corresponding to $\rho_\text{c}$=16000 g.cm$^{-3}$.}
    \label{Fig.Mode_Frequency_with_Rhoc}
\end{figure}

In Sections \ref{Sec.Interpolation_Results_of_Full_Grid} and \ref{Sec.Interpolation_Method_with_different_order_resolutions}, we noted that all measurements of the interpolation errors are significantly larger for the evolutionary track with $M$=1.175 M$_\odot$ (brown dashed line in Figure \ref{Fig.Interpolated_Errors_for_the_Full_Grid}). Here, we investigate this case further, to understand the origin of the larger errors and the implications that this extreme case may have in the construction of future model grids. To further characterize this anomaly, we address the results from the perspective of a fixed radial order and multiple evolutionary tracks, corresponding to the different masses in the grid. To that end, we compute the average offset in a way similar to that presented in Equation \ref{Eq.Average_offset_definition}, but with the average over radial orders replaced by an average over masses for the chosen radial order.
Figure \ref{Fig.Interpolation_in_rhoc_for_different_masses} presents the frequency interpolation errors for dipolar modes of radial order 11 as a function of the age proxies $\tau_\text{scaled}$ and $\rho_\text{c}$. 
For the mass track with $M$=1.175 M$_\odot$, the interpolation errors are not only about one order of magnitude larger than the typical errors found for the other masses, but they are also systematic, being always positive and always significant along the full evolutionary track.
This problem is not restricted to one particular dipole mode. In fact, all modes associated with the mass track with $M$=1.175 M$_\odot$ present similar maximum errors, that can be as high as $7.7\times10^0$ $\mu$Hz, and average offsets of the order of $1.7\times10^0$ $\mu$Hz.
This can be illustrated by focusing on a specific evolution point (e.g. at fixed $\rho_\text{c}$) and looking at how the interpolation error for a given radial order varies with mass. This is illustrated in Figure \ref{Fig.Mode_Frequency_with_Rhoc}, where we find a discontinuity between the evolutionary tracks with $M$=1.175 and 1.180 M$_\odot$.

\begin{figure}
    \resizebox{\hsize}{!}{\includegraphics{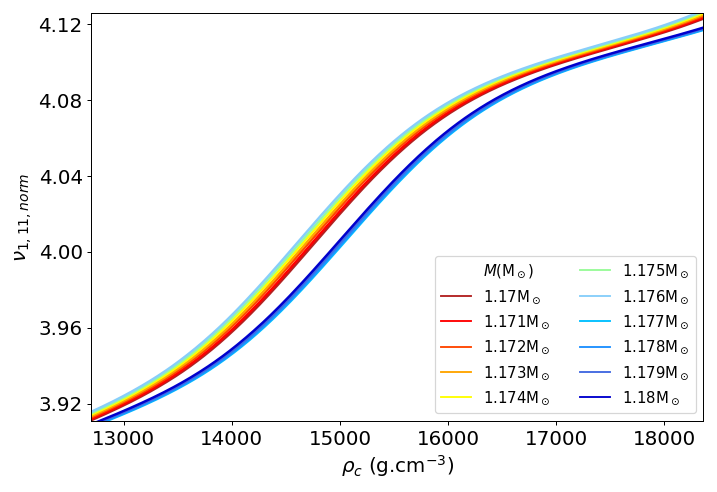}}
    \caption{Evolution of the mode frequency $\nu_\text{1,11,norm}$ along an avoided crossing taking place at central densities between $\rho_\text{c}$=13000 and 18000 g.cm$^{-3}$, for the models of the stellar subgrid with $f_\text{ov}=0.01$. The transition mass, between radiative and convective cores in the MS, is 1.176 M$_\odot$ for this choice of $f_\text{ov}$.}
    \label{Fig.Frequency_Rhoc_Curves_of_Transitional_Grid_0010}
\end{figure}

\begin{figure}
    \resizebox{\hsize}{!}{\includegraphics{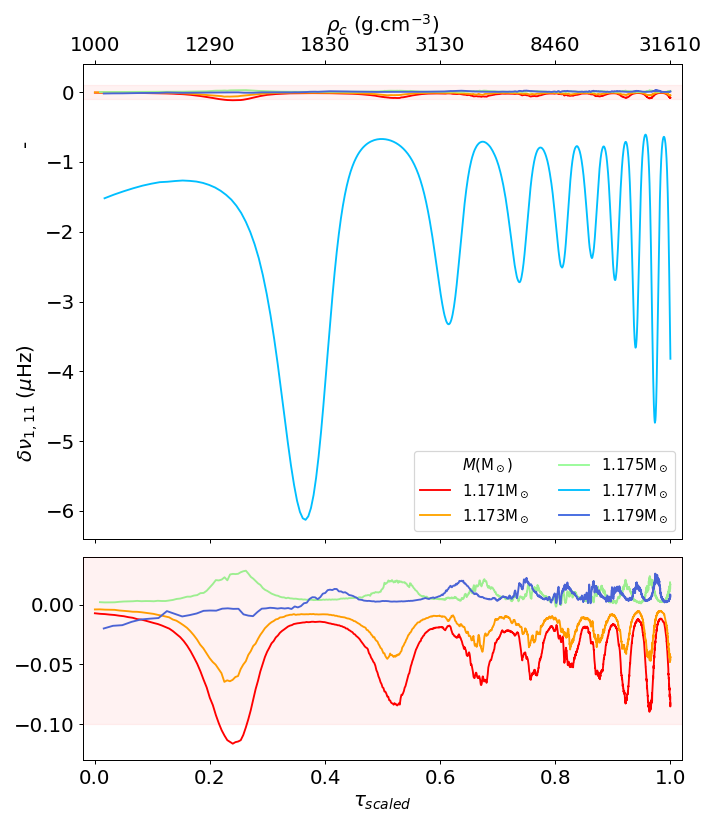}}
    \caption{Interpolation errors presented as a function of $\tau_\text{scaled}$ (bottom x-axis) and $\rho_\text{c}$ (top x-axis), that resulted from the application of the full interpolation algorithm (cubic splines along tracks and linear across tracks). Results are shown for all $\nu_{1,11}$ mode frequencies of the subgrid of stellar models with $f_\text{ov}=0.01$, when $\rho_\text{c}$ is the age proxy for interpolation. The transition mass, between radiative and convective cores in the MS, is 1.176 M$_\odot$ for this choice of $f_\text{ov}$. The bottom subplot presents a zoom in on the results of interpolation of the top subplot for $\delta\nu_{1,11}$ between $-$0.14 and 0.04 $\mu$Hz.}
    \label{Fig.Interpolation_in_rhoc_for_different_masses_of_transitional_grid_fov0010}
\end{figure}

To further understand the anomaly, we consider a smaller and denser subgrid purposely built to further explore this problem. This denser subgrid considers the exact same physics as the main grid described in Section \ref{Sec.Characterization_of_the_Method}, only with different mass inputs (that vary between 1.170 and 1.180 M$_\odot$, in steps of 0.001 M$_\odot$). 
Figure \ref{Fig.Frequency_Rhoc_Curves_of_Transitional_Grid_0010} shows the evolution of $\nu_{1,11,\text{norm}}$ with $\rho_\text{c}$ for the different mass tracks of such subgrid, zoomed in on a middle stage of the subgiant phase (in the interval $\rho_\text{c}$=[13000, 18000] g.cm$^{-3}$, or equivalently, $\tau_\text{scaled}$=[0.890, 0.925]). 
We can see that increasing the grid resolution in mass step worsens the problem, since the curves describing the evolution of the mode frequency along the evolutionary tracks with masses 1.176 and 1.177 M$_\odot$ are further apart than those corresponding to the masses of $M$=1.175 and 1.180 M$_\odot$.
Thus, the problem presented in Figure \ref{Fig.Interpolation_in_rhoc_for_different_masses} cannot be solved by increasing the grid resolution in a specific region of the parameter space of the grid. This is more prominently displayed in Figure \ref{Fig.Interpolation_in_rhoc_for_different_masses_of_transitional_grid_fov0010}, where the results of applying the full interpolation algorithm to this denser subgrid, namely to the track with $M$=1.177 M$_\odot$, show no clear improvement. 

Considering the dependence of the avoided crossings on the core conditions (cf. Section \ref{Sec.Impact_of_Mixed_Modes_in_Subgiants}), we conjecture that the discontinuity discussed above could be a result of different core properties associated with stellar models with different masses, since these mass values are typically associated with a transition between stars that had either a radiative core or a convective core during their MS.
This conjecture is supported by the evolution of the mass of the convective core in our grid, during the corresponding MS stages.
For evolutionary tracks with $M\leq$1.175 M$_\odot$, the mass of the convective core is non zero at the beginning of the MS, but it quickly becomes zero, remaining zero until the end of the MS. However, for $M\geq$1.175 M$_\odot$, the convective core mass is positive for most of the MS stage. 
This demonstrates a clear relation between the structure of the star during its MS and the systematic interpolation errors observed during the subgiant phase.

To investigate further the relation between the core structure and the interpolation errors, and explore possible ways to mitigate the problem, we consider the impact of the adopted core physics, in particular, the overshoot.
In particular, we consider a new subgrid, with similar physics as the main grid described in Section \ref{Sec.Characterization_of_the_Method}, but where we turn off the convective overshoot, by setting $f_\text{ov}=0.00$.
Such a decrease in $f_\text{ov}$ also implies a change in the transitional mass at which stars begin to develop a convective core. Thus, we consider different mass inputs for this subgrid (varying between 1.180 and 1.190 M$_\odot$), as we find the transitional mass for the case with no overshoot to be located between 1.182 and 1.183 M$_\odot$.

\begin{figure}
    \resizebox{\hsize}{!}{\includegraphics{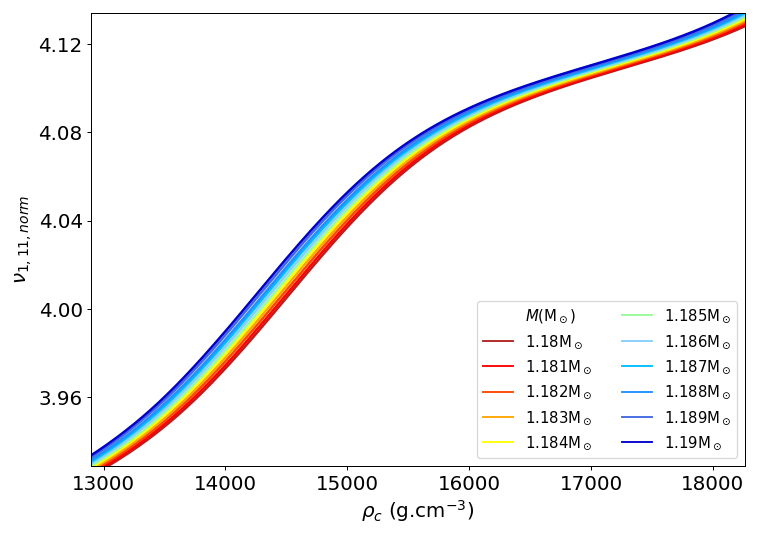}}
    \caption{Evolution of the mode frequency $\nu_\text{1,11,norm}$ along an avoided crossing taking place at central densities between $\rho_\text{c}$=13000 and 18000 g.cm$^{-3}$, for the models of the stellar subgrid with $f_\text{ov}=0.00$. The transition mass, between radiative and convective cores in the MS, is 1.182 M$_\odot$ for this choice of $f_\text{ov}$.}
    \label{Fig.Frequency_Rhoc_Curves_of_Transitional_Grid_0000}
\end{figure}

Figure \ref{Fig.Frequency_Rhoc_Curves_of_Transitional_Grid_0000} shows the evolution of the mode frequency $\nu_{1,11,\text{norm}}$ with central density $\rho_\text{c}$ for the case with no convective overshoot, zoomed in on the same stage of the subgiant evolution as in Figure \ref{Fig.Frequency_Rhoc_Curves_of_Transitional_Grid_0010}.
The discontinuity essentially disappears when $f_\text{ov}$ is set to zero. This shows that the cause of the discontinuity is not the appearance of a sustained convective core during the MS per se, but rather the abrupt change in the extent of the core across masses when an overshoot layer characterized by a fixed value of $f_\text{ov}$ is added at the core edge, using the current overshoot formulation. In fact, inspection of our models with overshoot shows that the abrupt change in core conditions when crossing the transition mass leads to a discontinuity in central temperature, $T_\text{c}$, at fixed $\rho_\text{c}$. This is then reflected in a jump in the frequencies of a given mode across the same mass transition.
To avoid this, it may be useful to consider as an age proxy the quantity $\rho_\text{c}^2\mu_\text{c}/T_\text{c}$ (cf. Equation \ref{Eq.Brunt_Vaisala_Frequency_in_Subgiant_Cores}), in place of $\rho_\text{c}$.

\begin{figure}
    \resizebox{\hsize}{!}{\includegraphics{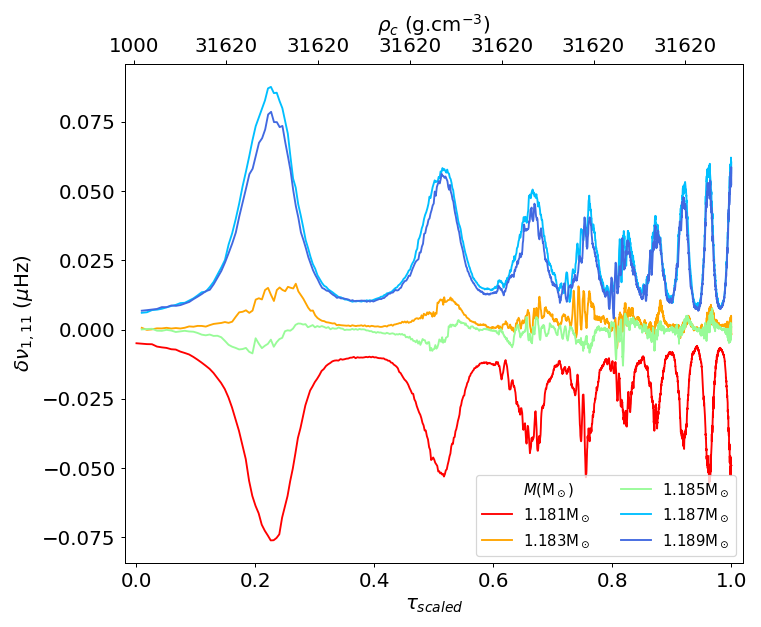}}
    \caption{Interpolation errors presented as a function of $\tau_\text{scaled}$ (bottom x-axis) and $\rho_\text{c}$ (top x-axis), that resulted from the application of the full interpolation algorithm (cubic splines along tracks and linear across tracks). Results are shown for all $\nu_{1,11}$ mode frequencies of the subgrid of stellar models with $f_\text{ov}=0.00$, when $\rho_\text{c}$ is the age proxy for interpolation. The transition mass, between radiative and convective cores in the MS, is 1.182 M$_\odot$ for this choice of $f_\text{ov}$.}
    \label{Fig.Interpolation_in_rhoc_for_different_masses_of_transitional_grid_fov0000}
\end{figure}

The results of applying the full interpolation method to this second subgrid are presented in Figure \ref{Fig.Interpolation_in_rhoc_for_different_masses_of_transitional_grid_fov0000}. In this case, the interpolation errors associated with the transitional mass (which corresponds to the interpolated mass-track with $M$=1.183 M$_\odot$) are no longer atypically large, with the maximum error being $8.9\times10^{-2}$ $\mu$Hz, well below the set accuracy limit of 0.1 $\mu$Hz.

Turning off $f_\text{ov}$ may not be the adequate way to mitigate the interpolation problem in this region of the grid, since this parameter describes important physics that affects the stellar structure of the subgiants. Still, these results show the importance of either ensuring that the grid is built such as to minimize the impact of abrupt structural changes on interpolation, or, in cases where there is no adequate way of doing that, accounting for the additional errors that may result from interpolation. In the specific case of core overshoot, a possible alternative is to consider a mass dependent overshoot, as suggested by \cite{Claret2018}, based on a study of eclipsing binaries. Such smooth variation in overshoot near the transition mass could potentially also mitigate the problem detected here.

\section{Conclusions}
\label{Sec.Conclusions_Article}

In this work, we performed a detailed study of classical interpolation algorithms applied to a grid of stellar models in the subgiant phase. Our primary goal was to verify whether these algorithms ensure the computation of the $\ell$=1 mode frequencies with interpolation errors within the typical observational errors ($\sim$0.1 $\mu$Hz).
The study was applied to a grid of 41 evolutionary tracks, varying only in stellar mass between 1.150 and 1.350 M$_\odot$, separated by steps of 0.005 M$_\odot$, chosen to cover the mass transition where their MS progenitors start to develop a sustained convective core. All other inputs were fixed, including the convective overshooting parameter, set to $f_\text{ov}$=0.01 in our initial grid. As a second goal, the work aimed at establishing the procedure, among those tested, that produces the smallest average offsets (for all modes, over evolution) and the smallest maximum absolute errors on the interpolated frequencies.

We tested different variations of the two-step interpolation algorithm, which considered various order methods (linear and cubic splines) and different age proxies (physical age, $\tau$, scaled age, $\tau_\text{scaled}$, and central density, $\rho_\text{c}$). We applied the algorithm to a grid with typical resolution in mass and age.
Our main conclusions are as follows: 1) the accuracy goal is not satisfied in any of the configurations considered, with the maximum interpolation error easily surpassing the reference accuracy limit of 0.1 $\mu$Hz by at least an order of magnitude; 2) the best results are obtained with the interpolation algorithm that considers cubic spline interpolation along evolutionary tracks, linear interpolation across evolutionary tracks, and $\rho_\text{c}$ as the age proxy. This interpolation algorithm yielded average offsets within twice the reference limit, but maximum errors up to an order of magnitude higher than that. 
These results show that the identification of an age proxy that is fully appropriate for the interpolation of mode frequencies in the presence of avoided crossings remains an open problem.

We also varied the density of models in the grid, starting with about 1500 models per evolutionary track to be used in interpolation, and decreasing such number progressively to 1/6 of that value, and investigated how the grid resolution influences the interpolation results.
We established that reducing the number of models per evolutionary track to 1/3 of the initial value ($\sim$500 models) does not significantly change the results, with the average offset remaining within twice the accuracy limit.
This seems to establish a minimal number of models per subgiant evolutionary track within the grid paradigm considered here, allowing future works to focus on larger ranges of stellar masses when building stellar grids, and additional input parameters, such as metallicity, [Fe/H], and convective overshoot, $f_\text{ov}$.
Such a grid could be used, e.g. for a reevaluation of the studies performed so far on Kepler subgiants data.
Nevertheless, we emphasize that, in all cases, the maximum errors across the grid are well above the established goal, and that regions in the grid in which the stellar structure varies abruptly are subject to even larger and systematic interpolation errors.

The work presented here is important in the context of space missions collecting asteroseismic data to characterize stars, such as PLATO, whose pipelines may rely on the use of stellar models. Our work indicates that, in the subgiant phase, state-of-the-art interpolation approaches in typical stellar grids will not enable the exploitation to their full potential of the exquisite seismic data that will be acquired. This, in turn, points to the need of developing a new generation of interpolation algorithms that may mitigate the problems identified in this work.

\begin{acknowledgements}
      This work was supported by Fundação para a Ciência e a Tecnologia (FCT/MCTES) through national funds by grants UIDB/04434/2020 (DOI: 10.54499/UIDB/04434/2020), UIDP/04434/2020 (DOI: 10.54499/UIDP/04434/2020), 2022.06962.PTDC, and 2022.03993.PTDC (DOI:10.54499/2022.03993.PTDC). MC is supported by FCT through a grant with reference 2020.07530.BD. MSC and TLC are supported by FCT through work contracts CEECIND/02619/2017 and CEECIND/00476/2018, respectively. SD acknowledges support from the project BEAMING ANR-18-CE31-0001 of the French National Research Agency (ANR) and from the Centre National d’Études Spatiales (CNES).
\end{acknowledgements}



\bibliographystyle{aa}
\bibliography{example} 


\begin{appendix}

\section{Interpolation errors resulting from the full interpolation algorithm}
\label{AppA:Errors_of_Interpolated_Mass_Tracks_of_Grid}

The following table of tables presents the average offset ($\mu$) and maximum error ($m$) (in $\mu$Hz) from the full interpolation method (that considers cubic splines along evolutionary tracks and the indicated interpolation method across evolutionary tracks), for all $\ell=1$ modes obtained for the interpolated tracks of the initial grid, with the interpolated mass value indicated on the top-left cell of each table.

\begin{table}
    \centering
    \caption{Average offset ($\mu$) and maximum errors ($m$) from the full interpolation algorithm (that considers cubic splines along tracks and the indicated interpolation method in each column across tracks). The results are for all $\ell$=1 modes associated with the missing evolutionary tracks of the subset grid, that appear in the initial grid (i.e. those whose third decimal place of the associated mass value is 5), determined from interpolation using adjacent evolutionary tracks, for each of the age proxies indicated.}
    \begin{tabular}{cccc}
    \hline \noalign{\smallskip}
    $M$=1.155 M$_\odot$ & Errors & Linear & Cubic Splines \\ 
    \noalign{\smallskip} \hline \noalign{\smallskip}
    \multirow{2}{4em}{$\tau$} & $\mu$ ($\mu$Hz) & $1.2\times10^1$ & --- \\
    & $m$ ($\mu$Hz) & $3.6\times10^1$ & --- \\
    \noalign{\smallskip} \hline \noalign{\smallskip}
    \multirow{2}{4em}{$\tau_\text{scaled}$} & $\mu$ ($\mu$Hz) & $4.2\times10^{-2}$ & --- \\
    & $m$ ($\mu$Hz) & $3.8\times10^{-1}$ & --- \\
    \noalign{\smallskip} \hline \noalign{\smallskip}
    \multirow{2}{4em}{$\rho_c$} & $\mu$ ($\mu$Hz) & $4.0\times10^{-2}$ & --- \\
    & $m$ ($\mu$Hz) & $4.5\times10^{-1}$ & --- \\
    \noalign{\smallskip} \hline
    \end{tabular}
\end{table}

\begin{table}
    \centering
    \begin{tabular}{cccc}
    \hline \noalign{\smallskip}
    $M$=1.165 M$_\odot$ & Errors & Linear & Cubic Splines \\ 
    \noalign{\smallskip} \hline \noalign{\smallskip}
    \multirow{2}{4em}{$\tau$} & $\mu$ ($\mu$Hz) & $1.2\times10^1$ & $1.1\times10^1$ \\
    & $m$ ($\mu$Hz) & $3.5\times10^1$ & $3.0\times10^1$ \\
    \noalign{\smallskip} \hline \noalign{\smallskip}
    \multirow{2}{4em}{$\tau_\text{scaled}$} & $\mu$ ($\mu$Hz) & $2.7\times10^{-2}$ & $1.5\times10^0$ \\
    & $m$ ($\mu$Hz) & $2.8\times10^{-1}$ & $3.5\times10^0$ \\
    \noalign{\smallskip} \hline \noalign{\smallskip}
    \multirow{2}{4em}{$\rho_c$} & $\mu$ ($\mu$Hz) & $3.4\times10^{-2}$ & $2.1\times10^{-1}$ \\
    & $m$ ($\mu$Hz) & $3.8\times10^{-1}$ & $9.5\times10^{-1}$ \\
    \noalign{\smallskip} \hline
    \end{tabular}
\end{table}

\begin{table}
    \centering
    \begin{tabular}{cccc}
    \hline \noalign{\smallskip}
    $M$=1.175 M$_\odot$ & Errors & Linear & Cubic Splines \\
    \noalign{\smallskip} \hline \noalign{\smallskip}
    \multirow{2}{4em}{$\tau$} & $\mu$ ($\mu$Hz) & $1.2\times10^1$ & $1.1\times10^1$ \\
    & $m$ ($\mu$Hz) & $3.4\times10^1$ & $2.9\times10^1$ \\
    \noalign{\smallskip} \hline \noalign{\smallskip}
    \multirow{2}{4em}{$\tau_\text{scaled}$} & $\mu$ ($\mu$Hz) & $1.2\times10^1$ & $1.2\times10^1$ \\
    & $m$ ($\mu$Hz) & $2.7\times10^1$ & $2.7\times10^1$ \\
    \noalign{\smallskip} \hline \noalign{\smallskip}
    \multirow{2}{4em}{$\rho_c$} & $\mu$ ($\mu$Hz) & $1.7\times10^0$ & $1.7\times10^0$ \\
    & $m$ ($\mu$Hz) & $7.7\times10^0$ & $7.5\times10^0$ \\
    \noalign{\smallskip} \hline
    \end{tabular}
\end{table}

\begin{table}
    \centering
    \begin{tabular}{cccc}
    \hline \noalign{\smallskip}
    $M$=1.185 M$_\odot$ & Errors & Linear & Cubic Splines \\
    \noalign{\smallskip} \hline \noalign{\smallskip}
    \multirow{2}{4em}{$\tau$} & $\mu$ ($\mu$Hz) & $8.2\times10^0$ & $1.1\times10^1$ \\
    & $m$ ($\mu$Hz) & $3.1\times10^1$ & $3.0\times10^1$ \\
    \noalign{\smallskip} \hline \noalign{\smallskip}
    \multirow{2}{4em}{$\tau_\text{scaled}$} & $\mu$ ($\mu$Hz) & $9.4\times10^{-2}$ & $1.8\times10^{0}$ \\
    & $m$ ($\mu$Hz) & $7.0\times10^{-1}$ & $3.7\times10^0$ \\
    \noalign{\smallskip} \hline \noalign{\smallskip}
    \multirow{2}{4em}{$\rho_c$} & $\mu$ ($\mu$Hz) & $6.9\times10^{-2}$ & $2.2\times10^{-1}$ \\
    & $m$ ($\mu$Hz) & $4.5\times10^{-1}$ & $1.1\times10^0$ \\
    \noalign{\smallskip} \hline
    \end{tabular}
\end{table}

\begin{table}
    \centering
    \begin{tabular}{cccc}
    \hline \noalign{\smallskip}
    $M$=1.195 M$_\odot$ & Errors & Linear & Cubic Splines \\
    \noalign{\smallskip} \hline \noalign{\smallskip}
    \multirow{2}{4em}{$\tau$} & $\mu$ ($\mu$Hz) & $7.8\times10^0$ & $7.4\times10^0$ \\
    & $m$ ($\mu$Hz) & $2.9\times10^1$ & $3.0\times10^1$ \\
    \noalign{\smallskip} \hline \noalign{\smallskip}
    \multirow{2}{4em}{$\tau_\text{scaled}$} & $\mu$ ($\mu$Hz) & $2.1\times10^{-1}$ & $8.3\times10^{-2}$ \\
    & $m$ ($\mu$Hz) & $1.6\times10^0$ & $1.1\times10^1$ \\
    \noalign{\smallskip} \hline \noalign{\smallskip}
    \multirow{2}{4em}{$\rho_c$} & $\mu$ ($\mu$Hz) & $1.6\times10^{-1}$ & $1.6\times10^{-1}$ \\
    & $m$ ($\mu$Hz) & $1.2\times10^0$ & $1.1\times10^0$ \\
    \noalign{\smallskip} \hline
    \end{tabular}
\end{table}

\begin{table}
    \centering
    \begin{tabular}{cccc}
    \hline \noalign{\smallskip}
    $M$=1.205 M$_\odot$ & Errors & Linear & Cubic Splines \\ 
    \noalign{\smallskip} \hline \noalign{\smallskip}
    \multirow{2}{4em}{$\tau$} & $\mu$ ($\mu$Hz) & $7.5\times10^0$ & $6.9\times10^0$ \\
    & $m$ ($\mu$Hz) & $2.7\times10^1$ & $2.9\times10^1$ \\
    \noalign{\smallskip} \hline \noalign{\smallskip}
    \multirow{2}{4em}{$\tau_\text{scaled}$} & $\mu$ ($\mu$Hz) & $2.4\times10^{-1}$ & $1.3\times10^{-1}$ \\
    & $m$ ($\mu$Hz) & $1.7\times10^0$ & $1.3\times10^0$ \\
    \noalign{\smallskip} \hline \noalign{\smallskip}
    \multirow{2}{4em}{$\rho_c$} & $\mu$ ($\mu$Hz) & $1.2\times10^{-1}$ & $1.1\times10^{-1}$ \\
    & $m$ ($\mu$Hz) & $1.1\times10^0$  & $1.0\times10^0$ \\
    \noalign{\smallskip} \hline
    \end{tabular}
\end{table}

\begin{table}
    \centering
    \begin{tabular}{cccc}
    \hline \noalign{\smallskip}
    $M$=1.215 M$_\odot$ & Errors & Linear & Cubic Splines \\ 
    \noalign{\smallskip} \hline \noalign{\smallskip}
    \multirow{2}{4em}{$\tau$} & $\mu$ ($\mu$Hz) & $7.4\times10^0$ & $6.7\times10^0$ \\
    & $m$ ($\mu$Hz) & $2.5\times10^1$ & $2.9\times10^1$ \\ 
    \noalign{\smallskip} \hline \noalign{\smallskip}
    \multirow{2}{4em}{$\tau_\text{scaled}$} & $\mu$ ($\mu$Hz) & $1.4\times10^{-1}$ & $1.8\times10^{-1}$ \\
    & $m$ ($\mu$Hz) & $1.6\times10^0$ & $1.2\times10^0$ \\ 
    \noalign{\smallskip} \hline \noalign{\smallskip}
    \multirow{2}{4em}{$\rho_c$} & $\mu$ ($\mu$Hz) & $6.5\times10^{-2}$ & $7.2\times10^{-2}$ \\
    & $m$ ($\mu$Hz) & $6.0\times10^{-1}$ & $6.7\times10^{-1}$ \\ 
    \noalign{\smallskip} \hline
    \end{tabular}
\end{table}

\begin{table}
    \centering
    \begin{tabular}{cccc}
    \hline \noalign{\smallskip}
    $M$=1.225 M$_\odot$ & Errors & Linear & Cubic Splines \\ 
    \noalign{\smallskip} \hline \noalign{\smallskip}
    \multirow{2}{4em}{$\tau$} & $\mu$ ($\mu$Hz) & $7.3\times10^0$ & $6.4\times10^0$ \\
    & $m$ ($\mu$Hz) & $2.5\times10^1$ & $2.8\times10^1$ \\ 
    \noalign{\smallskip} \hline \noalign{\smallskip}
    \multirow{2}{4em}{$\tau_\text{scaled}$} & $\mu$ ($\mu$Hz) & $1.1\times10^{-1}$ & $1.3\times10^{-1}$ \\
    & $m$ ($\mu$Hz) & $1.4\times10^0$ & $7.4\times10^{-1}$ \\
    \noalign{\smallskip} \hline \noalign{\smallskip}
    \multirow{2}{4em}{$\rho_c$} & $\mu$ ($\mu$Hz) & $4.9\times10^{-2}$ & $6.2\times10^{-2}$ \\
    & $m$ ($\mu$Hz) & $4.9\times10^{-1}$ & $4.9\times10^{-1}$ \\ 
    \noalign{\smallskip} \hline
    \end{tabular}
\end{table}

\begin{table}
    \centering
    \begin{tabular}{cccc}
    \hline \noalign{\smallskip}
    $M$=1.235 M$_\odot$ & Errors & Linear & Cubic Splines \\ 
    \noalign{\smallskip} \hline \noalign{\smallskip}
    \multirow{2}{4em}{$\tau$} & $\mu$ ($\mu$Hz) & $7.3\times10^0$ & $6.4\times10^0$ \\
    & $m$ ($\mu$Hz) & $2.5\times10^1$ & $2.7\times10^1$ \\ 
    \noalign{\smallskip} \hline \noalign{\smallskip}
    \multirow{2}{4em}{$\tau_\text{scaled}$} & $\mu$ ($\mu$Hz) & $5.5\times10^{-1}$ & $5.3\times10^{-1}$ \\
    & $m$ ($\mu$Hz) & $4.5\times10^0$ & $3.9\times10^0$ \\
    \noalign{\smallskip} \hline \noalign{\smallskip}
    \multirow{2}{4em}{$\rho_c$} & $\mu$ ($\mu$Hz) & $1.4\times10^{-1}$ & $1.4\times10^{-1}$ \\
    & $m$ ($\mu$Hz) & $1.2\times10^0$ & $1.3\times10^0$ \\ 
    \noalign{\smallskip} \hline
    \end{tabular}
\end{table}

\begin{table}
    \centering
    \begin{tabular}{cccc}
    \hline \noalign{\smallskip}
    $M$=1.245 M$_\odot$ & Errors & Linear & Cubic Splines \\ 
    \noalign{\smallskip} \hline \noalign{\smallskip}
    \multirow{2}{4em}{$\tau$} & $\mu$ ($\mu$Hz) & $7.3\times10^0$ & $6.1\times10^0$ \\
    & $m$ ($\mu$Hz) & $2.6\times10^1$ & $2.7\times10^1$ \\ 
    \noalign{\smallskip} \hline \noalign{\smallskip}
    \multirow{2}{4em}{$\tau_\text{scaled}$} & $\mu$ ($\mu$Hz) & $5.6\times10^{-1}$ & $5.7\times10^{-1}$ \\
    & $m$ ($\mu$Hz) & $3.6\times10^{0}$ & $3.3\times10^0$ \\ 
    \noalign{\smallskip} \hline \noalign{\smallskip}
    \multirow{2}{4em}{$\rho_c$} & $\mu$ ($\mu$Hz) & $1.4\times10^{-1}$ & $1.5\times10^{-1}$ \\
    & $m$ ($\mu$Hz) & $1.2\times10^0$ & $1.3\times10^0$ \\
    \noalign{\smallskip} \hline
    \end{tabular}
\end{table}

\begin{table}
    \centering
    \begin{tabular}{cccc}
    \hline \noalign{\smallskip}
    $M$=1.255 M$_\odot$ & Errors & Linear & Cubic Splines \\ 
    \noalign{\smallskip} \hline \noalign{\smallskip}
    \multirow{2}{4em}{$\tau$} & $\mu$ ($\mu$Hz) & $7.2\times10^0$ & $6.2\times10^0$ \\
    & $m$ ($\mu$Hz) & $2.6\times10^1$ & $2.7\times10^1$ \\ 
    \noalign{\smallskip} \hline \noalign{\smallskip}
    \multirow{2}{4em}{$\tau_\text{scaled}$} & $\mu$ ($\mu$Hz) & $1.5\times10^{-1}$ & $1.1\times10^{-1}$ \\
    & $m$ ($\mu$Hz) & $1.7\times10^0$ & $9.9\times10^{-1}$ \\ 
    \noalign{\smallskip} \hline \noalign{\smallskip}
    \multirow{2}{4em}{$\rho_c$} & $\mu$ ($\mu$Hz) & $2.7\times10^{-2}$ & $1.8\times10^{-2}$ \\
    & $m$ ($\mu$Hz) & $3.1\times10^{-1}$ & $2.5\times10^{-1}$ \\
    \noalign{\smallskip} \hline
    \end{tabular}
\end{table}

\begin{table}
    \centering
    \begin{tabular}{cccc}
    \hline \noalign{\smallskip}
    $M$=1.265 M$_\odot$ & Errors & Linear & Cubic Splines \\ 
    \noalign{\smallskip} \hline \noalign{\smallskip}
    \multirow{2}{4em}{$\tau$} & $\mu$ ($\mu$Hz) & $7.1\times10^0$ & $6.1\times10^0$ \\
    & $m$ ($\mu$Hz) & $2.6\times10^1$ & $2.6\times10^1$ \\ 
    \noalign{\smallskip} \hline \noalign{\smallskip}
    \multirow{2}{4em}{$\tau_\text{scaled}$} & $\mu$ ($\mu$Hz) & $9.1\times10^{-1}$ & $9.4\times10^{-1}$ \\
    & $m$ ($\mu$Hz) & $6.1\times10^0$ & $5.7\times10^0$ \\
    \noalign{\smallskip} \hline \noalign{\smallskip}
    \multirow{2}{4em}{$\rho_c$} & $\mu$ ($\mu$Hz) & $2.0\times10^{-1}$ & $2.1\times10^{-1}$ \\
    & $m$ ($\mu$Hz) & $1.7\times10^0$ & $1.8\times10^0$ \\ 
    \noalign{\smallskip} \hline
    \end{tabular}
\end{table}

\begin{table}
    \centering
    \begin{tabular}{cccc}
    \hline \noalign{\smallskip}
    $M$=1.275 M$_\odot$ & Errors & Linear & Cubic Splines \\ 
    \noalign{\smallskip} \hline \noalign{\smallskip}
    \multirow{2}{4em}{$\tau$} & $\mu$ ($\mu$Hz) & $7.2\times10^0$ & $6.0\times10^0$ \\
    & $m$ ($\mu$Hz) & $2.6\times10^1$ & $2.5\times10^1$ \\ 
    \noalign{\smallskip} \hline \noalign{\smallskip}
    \multirow{2}{4em}{$\tau_\text{scaled}$} & $\mu$ ($\mu$Hz) & $3.0\times10^{-1}$ & $2.1\times10^{-1}$ \\
    & $m$ ($\mu$Hz) & $3.3\times10^0$ & $2.4\times10^0$ \\
    \noalign{\smallskip} \hline \noalign{\smallskip}
    \multirow{2}{4em}{$\rho_c$} & $\mu$ ($\mu$Hz) & $4.4\times10^{-2}$ & $3.5\times10^{-2}$ \\
    & $m$ ($\mu$Hz) & $4.8\times10^{-1}$ & $2.7\times10^{-1}$ \\ 
    \noalign{\smallskip} \hline
    \end{tabular}
\end{table}

\begin{table}
    \centering
    \begin{tabular}{cccc}
    \hline \noalign{\smallskip}
    $M$=1.285 M$_\odot$ & Errors & Linear & Cubic Splines \\ 
    \noalign{\smallskip} \hline \noalign{\smallskip}
    \multirow{2}{4em}{$\tau$} & $\mu$ ($\mu$Hz) & $7.3\times10^0$ & $6.0\times10^0$ \\
    & $m$ ($\mu$Hz) & $2.7\times10^1$ & $2.3\times10^1$ \\ 
    \noalign{\smallskip} \hline \noalign{\smallskip}
    \multirow{2}{4em}{$\tau_\text{scaled}$} & $\mu$ ($\mu$Hz) & $7.0\times10^{-1}$ & $7.1\times10^{-1}$ \\
    & $m$ ($\mu$Hz) & $5.4\times10^0$ & $4.7\times10^0$ \\ 
    \noalign{\smallskip} \hline \noalign{\smallskip}
    \multirow{2}{4em}{$\rho_c$} & $\mu$ ($\mu$Hz) & $1.6\times10^{-1}$ & $1.7\times10^{-1}$ \\
    & $m$ ($\mu$Hz) & $1.1\times10^0$ & $1.2\times10^0$ \\ 
    \noalign{\smallskip} \hline
    \end{tabular}
\end{table}

\begin{table}
    \centering
    \begin{tabular}{cccc}
    \hline \noalign{\smallskip}
    $M$=1.295 M$_\odot$ & Errors & Linear & Cubic Splines \\
    \noalign{\smallskip} \hline \noalign{\smallskip}
    \multirow{2}{4em}{$\tau$} & $\mu$ ($\mu$Hz) & $7.2\times10^0$ & $6.0\times10^0$ \\
    & $m$ ($\mu$Hz) & $2.7\times10^1$ & $2.3\times10^1$ \\
    \noalign{\smallskip} \hline \noalign{\smallskip}
    \multirow{2}{4em}{$\tau_\text{scaled}$} & $\mu$ ($\mu$Hz) & $5.4\times10^{-1}$ & $5.6\times10^{-1}$ \\
    & $m$ ($\mu$Hz) & $4.2\times10^0$ & $3.5\times10^0$ \\
    \noalign{\smallskip} \hline \noalign{\smallskip}
    \multirow{2}{4em}{$\rho_c$} & $\mu$ ($\mu$Hz) & $1.2\times10^{-1}$ & $1.2\times10^{-1}$ \\
    & $m$ ($\mu$Hz) & $8.8\times10^{-1}$ & $8.8\times10^{-1}$ \\
    \noalign{\smallskip} \hline 
    \end{tabular}
\end{table}

\begin{table}
    \centering
    \begin{tabular}{cccc}
    \hline \noalign{\smallskip}
    $M$=1.305 M$_\odot$ & Errors & Linear & Cubic Splines \\ 
    \noalign{\smallskip} \hline \noalign{\smallskip}
    \multirow{2}{4em}{$\tau$} & $\mu$ ($\mu$Hz) & $7.3\times10^0$ & $6.1\times10^0$ \\
    & $m$ ($\mu$Hz) & $2.7\times10^1$ & $2.2\times10^1$ \\ 
    \noalign{\smallskip} \hline \noalign{\smallskip}
    \multirow{2}{4em}{$\tau_\text{scaled}$} & $\mu$ ($\mu$Hz) & $6.6\times10^{-1}$ & $6.0\times10^{-1}$ \\
    & $m$ ($\mu$Hz) & $5.8\times10^0$ & $5.0\times10^0$ \\ 
    \noalign{\smallskip} \hline \noalign{\smallskip}
    \multirow{2}{4em}{$\rho_c$} & $\mu$ ($\mu$Hz) & $1.1\times10^{-1}$ & $9.7\times10^{-2}$ \\
    & $m$ ($\mu$Hz) & $7.4\times10^{-1}$ & $7.2\times10^{-1}$ \\ 
    \noalign{\smallskip} \hline
    \end{tabular}
\end{table}

\begin{table}
    \centering
    \begin{tabular}{cccc}
    \hline \noalign{\smallskip}
    $M$=1.315 M$_\odot$ & Errors & Linear & Cubic Splines \\ 
    \noalign{\smallskip} \hline \noalign{\smallskip}
    \multirow{2}{4em}{$\tau$} & $\mu$ ($\mu$Hz) & $7.3\times10^0$ & $6.1\times10^0$ \\
    & $m$ ($\mu$Hz) & $2.6\times10^1$ & $2.2\times10^1$ \\ 
    \noalign{\smallskip} \hline \noalign{\smallskip}
    \multirow{2}{4em}{$\tau_\text{scaled}$} & $\mu$ ($\mu$Hz) & $2.5\times10^{-1}$ & $1.7\times10^{-1}$ \\
    & $m$ ($\mu$Hz) & $2.9\times10^0$ & $1.8\times10^0$ \\
    \noalign{\smallskip} \hline \noalign{\smallskip}
    \multirow{2}{4em}{$\rho_c$} & $\mu$ ($\mu$Hz) & $4.0\times10^{-2}$ & $3.3\times10^{-2}$ \\
    & $m$ ($\mu$Hz) & $2.8\times10^{-1}$ & $2.4\times10^{-1}$ \\
    \noalign{\smallskip} \hline
    \end{tabular}
\end{table}

\begin{table}
    \centering
    \begin{tabular}{cccc}
    \hline \noalign{\smallskip}
    $M$=1.325 M$_\odot$ & Errors & Linear & Cubic Splines \\ 
    \noalign{\smallskip} \hline \noalign{\smallskip}
    \multirow{2}{4em}{$\tau$} & $\mu$ ($\mu$Hz) & $7.4\times10^0$ & $6.3\times10^0$ \\
    & $m$ ($\mu$Hz) & $2.5\times10^1$ & $2.2\times10^1$ \\ 
    \noalign{\smallskip} \hline \noalign{\smallskip}
    \multirow{2}{4em}{$\tau_\text{scaled}$} & $\mu$ ($\mu$Hz) & $5.1\times10^{-1}$ & $3.6\times10^{-1}$ \\
    & $m$ ($\mu$Hz) & $5.0\times10^0$ & $4.0\times10^0$ \\ 
    \noalign{\smallskip} \hline \noalign{\smallskip}
    \multirow{2}{4em}{$\rho_c$} & $\mu$ ($\mu$Hz) & $1.1\times10^{-1}$ & $9.4\times10^{-2}$ \\
    & $m$ ($\mu$Hz) & $6.8\times10^{-1}$ & $6.1\times10^{-1}$ \\ 
    \noalign{\smallskip} \hline
    \end{tabular}
\end{table}

\begin{table}
    \centering
    \begin{tabular}{cccc}
    \hline \noalign{\smallskip}
    $M$=1.335 M$_\odot$ & Errors & Linear & Cubic Splines \\ 
    \noalign{\smallskip} \hline \noalign{\smallskip}
    \multirow{2}{4em}{$\tau$} & $\mu$ ($\mu$Hz) & $7.5\times10^0$ & $6.4\times10^0$ \\
    & $m$ ($\mu$Hz) & $2.3\times10^1$ & $2.2\times10^1$ \\ 
    \noalign{\smallskip} \hline \noalign{\smallskip}
    \multirow{2}{4em}{$\tau_\text{scaled}$} & $\mu$ ($\mu$Hz) & $8.6\times10^{-1}$ & $8.1\times10^{-1}$ \\
    & $m$ ($\mu$Hz) & $7.0\times10^0$ & $6.2\times10^0$ \\
    \noalign{\smallskip} \hline \noalign{\smallskip}
    \multirow{2}{4em}{$\rho_c$} & $\mu$ ($\mu$Hz) & $1.5\times10^{-1}$ & $1.4\times10^{-1}$  \\
    & $m$ ($\mu$Hz) & $9.6\times10^{-1}$ & $9.5\times10^{-1}$ \\
    \noalign{\smallskip} \hline
    \end{tabular}
\end{table}

\begin{table}
    \centering
    \begin{tabular}{cccc}
    \hline \noalign{\smallskip}
    $M$=1.345 M$_\odot$ & Errors & Linear & Cubic Splines \\ 
    \noalign{\smallskip} \hline \noalign{\smallskip}
    \multirow{2}{4em}{$\tau$} & $\mu$ ($\mu$Hz) & $7.7\times10^0$ & --- \\
    & $m$ ($\mu$Hz) & $2.3\times10^1$ & --- \\ 
    \noalign{\smallskip} \hline \noalign{\smallskip}
    \multirow{2}{4em}{$\tau_\text{scaled}$} & $\mu$ ($\mu$Hz) & $4.1\times10^{-1}$ & --- \\
    & $m$ ($\mu$Hz) & $4.0\times10^0$ & --- \\ 
    \noalign{\smallskip} \hline \noalign{\smallskip}
    \multirow{2}{4em}{$\rho_c$} & $\mu$ ($\mu$Hz) & $4.1\times10^{-2}$ & --- \\
    & $m$ ($\mu$Hz) & $4.5\times10^{-1}$ & --- \\ 
    \noalign{\smallskip} \hline
    \end{tabular}
\end{table}

\end{appendix}

\end{document}